\documentclass[acmtog,nonacm]{acmart}

\usepackage{makecell}
\usepackage{multirow}
\usepackage{soul}
\usepackage{subcaption}
\usepackage{enumitem}
\usepackage{boldline}
\usepackage{dblfloatfix} 
\AtBeginDocument{%
  \providecommand\BibTeX{{%
    \normalfont B\kern-0.5em{\scshape i\kern-0.25em b}\kern-0.8em\TeX}}}

\acmJournal{TOG}
\acmVolume{0}
\acmNumber{0}
\acmArticle{0}
\acmMonth{0}

\newcommand{\remove}[1]{}
\definecolor{darkblue}{rgb}{0,0,0.8}
\newcommand{\modified}[1]{#1}
\newcommand{\modifiedbis}[1]{#1}

\begin{document}

\title{Textured Mesh Quality Assessment: Large-Scale Dataset and Deep Learning-based Quality Metric}

\author{Yana Nehmé}
\email{yana.nehme@gmail.com}
\affiliation{
  \institution{Univ Lyon, INSA Lyon, CNRS, UCBL, LIRIS, UMR5205}
 \city{F-69621 Villeurbanne}
  \country{France}}

\author{Johanna Delanoy}
\email{johanna.delanoy@insa-lyon.fr}
\affiliation{
  \institution{Univ Lyon, INSA Lyon, CNRS, UCBL, LIRIS, UMR5205}
  \city{F-69621 Villeurbanne}
  \country{France}}

\author{Florent Dupont}
\email{Florent.Dupont@liris.cnrs.fr}
\affiliation{
  \institution{Univ Lyon, UCBL, CNRS, INSA Lyon, LIRIS, UMR5205}
  \city{F-69622 Villeurbanne}
  \country{France}}
  
\author{Jean-Philippe Farrugia}
\email{jean-philippe.farrugia@univ-lyon1.fr}
\affiliation{
  \institution{Univ Lyon, UCBL, CNRS, INSA Lyon, LIRIS, UMR5205}
  \city{F-69622 Villeurbanne}
  \country{France}}

\author{Patrick Le Callet}
\email{patrick.lecallet@univ-nantes.fr}
\affiliation{
  \institution{Nantes Université, École Centrale Nantes, CNRS, LS2N, UMR 6004}
  \city{F-44000 Nantes}
  \country{France}}

\author{Guillaume Lavoué}
\email{glavoue@liris.cnrs.fr}
\affiliation{
  \institution{Univ Lyon, Centrale Lyon, CNRS, INSA Lyon, UCBL, LIRIS, UMR5205, ENISE}
  \city{F-42023 Saint Etienne}
  \country{France}}

\renewcommand{\shortauthors}{Nehmé et al.}

\begin{abstract}
Over the past decade, 3D graphics have become highly detailed to mimic the real world, exploding their size and complexity. Certain applications and device constraints necessitate their simplification and/or lossy compression, which can degrade their visual quality.
Thus, to ensure the best Quality of Experience (QoE), it is important to evaluate the visual quality to accurately drive the compression and find the right compromise between visual quality and data size.
In this work, we focus on subjective and objective quality assessment of textured 3D meshes. 
We first establish a large-scale dataset, which includes 55 source models quantitatively characterized in terms of geometric, color, and semantic complexity, and corrupted by combinations of 5 types of compression-based distortions applied on the geometry, texture mapping and texture image of the meshes.  This dataset contains over 343k distorted stimuli.
We  propose an approach to select a challenging subset of 3000 stimuli for which we collected 148929 quality judgments from over 4500 participants in a large-scale crowdsourced subjective experiment. 
Leveraging our subject-rated dataset, a learning-based quality metric for 3D graphics was proposed.  Our metric demonstrates state-of-the-art results on our dataset of textured meshes and on a dataset of distorted meshes with vertex colors.
Finally, we present an application of our metric and dataset to explore the influence of distortion interactions and content characteristics on the perceived quality of compressed textured meshes.
\end{abstract}



\maketitle

\section{Introduction}
The use of 3D graphical data is growing for the general public with the proliferation of acquisition technologies (3D scanners,  360 \textdegree cameras, MRI, etc.), intuitive 3D modeling tools, 3D printers, and affordable virtual and mixed reality Head-Mounted Displays -HMD- (Oculus Rift, HTC Vive, Microsoft HoloLens, etc.). 
All of these technologies make the size and complexity of 3D data explode. The resulting 3D scenes are huge and extremely detailed: they may contain several million geometric primitives, associated with a wide range of appearance attributes, intended to reproduce a realistic material appearance.

Extended reality -XR- (i.e. Augmented and Virtual Reality AR/VR) is seen as the next potential computing platform.
The great advantage of XR technologies is that they provide 6 Degrees of Freedom (6DoF) allowing realistic interactions and a high level of immersion. 
However, the visualization and interaction in XR of large and complex 3D scenes remains an unsolved issue to date due to two major challenges: (1) the complexity of a 3D scene that can be displayed on a HMD is substantially smaller than that on a standard screen, (2) for networked applications, latency problems may occur when streaming the 3D scene on the client device.  This problem is growing as more online VR/AR applications consider 3D data stored on remote servers.

To adapt the complexity of 3D content for HMDs (notably for autonomous devices, such as the Occulus Quest 2) and to avoid latency due to transmission, simplification and compression are inevitable. These operations reduce the amount of data (reduce the Level of Details (LoD) and the size of 3D data) and by extension the costs in processing, storage, and transmission.  However, such operations are lossy and result in visual degradations that may affect the perceived quality of the 3D scene and, in turn, the user's Quality Of Experience (QoE). 
It is therefore essential to define measures to accurately assess the impact of these distortions in order to find the right compromise between visual quality and data size/LoD. For this purpose, quality assessment methodologies are required.

The perceptual quality can be assessed using subjective studies and objective metrics.  Objective metrics consist in algorithms designed to automatically predict the visual quality loss (i.e. the level of annoyance of visual artifacts). On the other hand, subjective studies, aka. user studies, involve inviting a group of participants to assess the visual quality of test data. These subjective experiments provide the most reliable way to create ground-truth datasets useful for understanding human psychological behavior (when perceiving multimedia content) as well as for benchmarking and tuning objective quality metrics.  

Public quality assessment metrics and datasets for 3D graphics are lacking, especially for meshes with color and/or texture attributes.  Indeed, existing datasets are rather small and have limited generalization ability, making them not challenging enough to test and benchmark quality metrics and insufficient to train and drive learning-based ones.

In this work, we produce a valuable large-scale quality assessment dataset of textured meshes, with more than 343k distorted meshes generated from the compression of 55 source models.
To ensure the generalization ability of our dataset, (1) we devised a set of measures to quantitatively characterize the source models in order to avoid biases related to the selection of these models.  
(2) We involved mixed compression-based distortions of different natures, such as texture compression, geometry quantization, UV map quantization, LoD and texture sub-sampling.
(3) We proposed an approach for the selection of a challenging subset of 3000 stimuli that was subject-rated in a large-scale crowdsourced experiment.

Leveraging our annotated dataset, we propose an image-based deep learning quality metric for 3D graphics, called Graphics-LPIPS \modified{(Learned Perceptual Image Patch Similarity)}. It operates on patches of rendered stimuli that are fed to a convolutional neural network with learning weights on top to extract features.  Features are then fused and pooled to predict the quality of the patch. 
The global quality score of the stimulus is obtained by averaging the quality of local patches.
Our metric outperforms other image quality metrics in terms of correlation and classification ability on our dataset of textured meshes and on an existing dataset of meshes with vertex colors.

Finally, our metric allowed us to annotate the remaining stimuli in our dataset to subsequently explore the influence of distortion interactions and content characteristics on the perceived quality of 3D graphics.\\

The main contributions of our work are as follows:
\begin{itemize}
	\item We provide the largest dataset of textured meshes with over 343k stimuli generated from 55 source models quantitatively characterized in terms of geometric, color, and semantic complexity to ensure their diversity.  The dataset covers a wide range of compression-based distortions applied with different strengths. The database can be used to train no-reference quality metrics and develop rate-distortion models for meshes.
	\item From the established dataset, we carefully selected a challenging subset of 3000 stimuli that we annotated in a large-scale subjective experiment in crowdsourcing involving over 4500 participants. To the best of our knowledge, it is the largest quality assessment dataset of textured meshes associated with subjective scores to date. This database is valuable for training and benchmarking quality metrics.
	\item We propose an image-based metric, named Graphics-LPIPS,  for assessing the quality of rendered 3D graphics. It employs convolutional neural networks. Our metric demonstrates state-of-the-art results on two different datasets.
	\item Leveraging our whole dataset and metric, we provide an in-depth analysis on the effect of each distortion and their combinations on the perceived quality of textured meshes. We also evaluate the influence of the geometric and color complexity of the model on the perception of distortions.
\end{itemize}
The datasets and the source code of the metric are publicly available \footnote{https://github.com/MEPP-team/Graphics-LPIPS}.\\

The rest of this paper is organized as follows: 
Section \ref{sec:relatedwork} reviews previous work on subjective and objective quality assessment of 3D graphics. 
Section \ref{sec:TexturedDBGeneration} describes our dataset and the set of measures we propose for 3D content characterization. 
Section \ref{sec:TextDB_Exp} details the subjective experiment and the process adopted to select the subset of test stimuli.
We describe in section \ref{sec:Graphics-LPIPS} our learning-based quality metric and evaluate its performance.
Section \ref{sec:applicationGraphics-LPIPS} presents an application of the metric and the dataset. 
Concluding remarks and perspectives are provided in section \ref{sec:conclusion}.



\section{Related work}\label{sec:relatedwork}
In this section, we review previous work on subjective and objective quality assessment of graphical 3D content. We provide an overview of existing datasets and metrics for predicting the visual impact of distortions applied on such data. 
Note that, 3D data can be represented in different ways (3D meshes, point clouds, voxels), with and without appearance attributes (color, texture, material, etc.). We are specifically interested in 3D meshes and point clouds with color/texture attributes.

\subsection{Subjective quality assessment} \label{subsec:RelatedWork_subj}
Subjective quality tests involving 3D models were initially introduced on meshes, more precisely on geometry-only models, to assess the artifacts induced by the simplification, smoothing, watermarking and compression \cite{Watson2001, Corsini2007, Lavoue2009, Vasa2012, Torkhani2015, Vanhoey2017, Christaki2018}. 
Little work considered meshes with color attributes (vertex color or texture) \cite{Pan2005, Guo2016, Zerman2020a, Gutierrez2020, Nehme2020TAP, Nehme2021TVCG}.

Subjective quality experiments involving point clouds have become prevalent over the last six years.
The pioneering studies were conducted on colorless point cloud content~\cite{Alexiou2017b, Alexiou2017c, Javaheri2017b, Su2019a, Javaheri2019a}. 
Subsequently, the majority of the studies focused on evaluating the quality of colored point clouds 
~\cite{Perry2020a,  Liu2021, Torlig2018a, Yang2020b, Wu2021a, Alexiou2019c,  Zerman2019a,  Zerman2020a, Subramanyam2020a}.

The aforementioned works considered different subjective methodologies,  different ways to display the 3D models to participants (still images, animations,  interactive scenes) and different test equipment (2D screen, augmented reality and virtual reality headsets).

The experimental methodologies used were inspired by existing image/video subjective methodologies. They are mainly derived from single stimulus methods, in which participants see only one stimulus and rate its quality \cite{Corsini2007,Torkhani2015,Zerman2020a,Subramanyam2020a,Gutierrez2020,viola2022impact},   double stimulus methods in which participants rate the visual degradation after seeing the reference and distorted stimuli \cite{Watson2001, Lavoue2009,Torlig2018a, Su2019a, Perry2020a, Nehme2021TVCG}, 
and pairwise comparison methods in which participants choose the better quality stimulus from two stimuli presented to them \cite{Vasa2012, Guo2016, Vanhoey2017, Christaki2018,Alexiou2019c}.
Recently,  a couple of comprehensive/comparative studies \cite{Alexiou2017b,  Nehme2020TAP} evaluated the impact of the subjective methodologies on the obtained quality scores.  They found that the double stimulus method is the most suitable to assess the quality of 3D graphics.

Some researchers have implemented real-time interactive scenes, allowing participants to freely interact (rotate and zoom) with the 3D models being rated (real-time interactive inspection) 
\cite{Alexiou2017b, Torlig2018a,Mekuria2017a, Alexiou2017c, Alexiou2020b,Wu2021a,Gutierrez2020, Yang2020b,Subramanyam2020a}. 
These experiments are conducted on desktop devices as well as in immersive environments with varying Degrees of Freedom (DoF). 
Other researchers have controlled the viewpoints visualized by the participants to ensure the same user experience (passive inspection). They used 2D still images or predefined camera paths to generate videos of the models \cite{Rogowitz2001, Pan2005,  Guo2016, Zerman2020a, Nehme2021TVCG,Javaheri2017b, Su2019a,Zerman2019a}.  This approach avoids cognitive overload that can alter human judgments.
The effect of adopting different modes of inspection for subjective quality assessment is still unclear/an open question as very few comparisons have been made to date \cite{Torkhani2015, viola2022impact}.

Most of the reported experiments for both meshes and point clouds were conducted on desktop settings.  Only the studies presented in~\cite{Christaki2018, Alexiou2020b, Subramanyam2020a, Nehme2020TAP, Nehme2021TVCG, Wu2021a, viola2022impact} considered a VR environment, and those presented in \cite{Gutierrez2020,  Alexiou2017c} a AR environment.  An early attempt of 3D tele-immersion was reported in~\cite{Mekuria2017a}.
So far,  no work has been done to understand the impact of display devices on the perceived quality of 3D content.  

The experiments presented above were conducted in laboratories.  In recent years,  CrowdSourcing (CS) experiments have gained popularity,  as they are relatively fast and are therefore more practical for evaluating large-scale datasets.  
A recent study investigated whether a crowdsourcing test can achieve the accuracy of a laboratory test for the quality assessment of 3D graphics \cite{Nehme2021MMSP}.
The results showed that under controlled conditions and with a proper participant screening approach, a crowdsourcing experiment based on the double stimulus method can be as accurate as a laboratory experiment (based on the same methodology). 
Another crowdsourcing study evaluated the perception of compression distortions on point clouds \cite{Lazzarotto2021a}.

Several works presented above have publicly released their datasets. 
Table \ref{tab:ResultingDB} lists the publicly available quality assessment datasets for 3D content with color/texture attributes. 
\begin{table}[tbhp]
\centering
\caption{Publicly available quality assessment datasets for meshes and point clouds.}
\label{tab:ResultingDB}
\resizebox{1\linewidth}{!}{%
\begin{tabular}{l|lcc} 
\multicolumn{1}{c|}{Dataset} & \makecell{3D Repre-\\sentation} & Attributes  &\makecell{\# Stimuli\\rated}  \\ \midrule
\makecell[l]{LIRIS Textured Mesh\\\cite{Guo2016}} & Mesh & Texture maps & \makecell{$\bullet$ 100$\times$2 renderings\\$\bullet$ 36$\times$2 renderings} \\ \hline
\makecell[l]{3D Meshes with Vertex\\Colors~\cite{Nehme2021TVCG}} & Mesh & Vertex colors & 480 \\ \clineB{1-4}{3}
\makecell[l]{M-PCCD\\ \cite{Alexiou2019c}}& Point cloud  & Colored    & \makecell[c]{$\bullet$ 240\\$\bullet$ 40 \& 30}  \\ \hline
IRPC~\cite{Javaheri2019a} & Point cloud    &  \makecell[l]{$\bullet$ 2$\times$Colorless \\$\bullet$ Colored}     & \makecell[c]{$\bullet$ 54 \\$\bullet$ 54}   \\ \hline
WPC~\cite{Su2019a}  & Point cloud   & Colored   & 740   \\  \hline
\makecell[l]{VsenseVVDB\\\cite{Zerman2019a}} & Point cloud   & Colored   & 32   \\  \hline
\makecell[l]{VsenseVVDB2\\\cite{Zerman2020a}} & \makecell[l]{$\bullet$ Point cloud \\$\bullet$ Mesh}  & \makecell[l]{$\bullet$ Colored \\$\bullet$ Texture maps}&  \makecell[c]{$\bullet$ 136 \\$\bullet$ 28} \\  \hline
\makecell[l]{ICIP2020 \cite{Perry2020a}} & Point cloud   & Colored  & 96  \\  \hline
\makecell[l]{PointXR~\cite{Alexiou2020b}} & Point cloud  & Colored   & 40 \\ \hline
\makecell[l]{SJTU-PCQA~\cite{Yang2020b}} & Point cloud   & Colored  & 378 \\  \hline
\makecell[l]{SIAT-PCQD \cite{Wu2021a}}  & Point cloud  & Colored  & 340  \\ \hline
\makecell[l]{LB-PCCD\\\cite{Lazzarotto2021a}} & Point cloud  & Colored   & 105   \\ \hline
\makecell[l]{2DTV-VR-QoE\\\cite{viola2022impact}} & Point cloud  & Colored  & 72 \\ \hline
\makecell[l]{LS-PCQA \cite{Liu2021}\\Available after publication}  & Point cloud  & Colored  & \makecell{$\bullet$ 1320\\(MOS) \\$\bullet$ 22704\\(Pseudo-MOS)} \\ \clineB{1-4}{3}
Our dataset & Mesh  & Texture maps  & \makecell{$\bullet$ 3000\\(MOS) \\$\bullet$ 340750\\(Pseudo-MOS)}  \\
\bottomrule
\end{tabular}}
\end{table}
There is a lack of large-scale 3D content datasets, especially those for meshes with color attributes, either in the form of vertex colors or texture maps. 
Existing datasets are rather small: they contain only few hundreds distorted stimuli, which is not sufficient to drive deep learning metrics that rely on the richness and generality of datasets.
In this work, we produce the largest quality assessment dataset of textured meshes to date.  In total, more than 343k distorted meshes were generated, of which 3000 are associated with subjective Mean Opinion Scores (MOS) derived from a large-scale subjective experiment conducted in crowdsourcing. Quality scores of the remaining stimuli were predicted (Pseudo-MOS) using a proposed quality metric based on deep learning,  trained and tested on the subset of annotated stimuli. 
Our large dataset allowed us to analyze the impact of the distortions and model characteristics on the perceived quality of textured meshes.

\subsection{Objective quality assessment}
Simple geometric measures, such as Hausdorff distance \cite{Aspert2002}, Root Mean Squared (RMS) error, and Peak Signal-to-Noise Ratio (PSNR), are only weakly correlated with the human vision since they are based on pure geometric distances and ignore perceptual information \cite{Lavoue2016,Zerman2019a}. Therefore, many perceptually-driven visual quality metrics have been proposed for meshes and point clouds. 

The most popular of these metrics are based on top-down approaches. They treat the Human Visual System (HVS) as a black box and identify changes in content features induced by distortions to estimate perceived quality.  They are mostly full-reference and work in a similar way by first establishing the correspondence between the reference and degraded models, after which a set of local feature errors are computed locally (over a neighborhood around each point/vertex) and then pooled into a global quality score. Early metrics developed evaluate only geometric distortions such as MSDM2 \cite{Lavoue2011}, DAME \cite{Vasa2012},  FMPD \cite{Wang2012},  
and TPDM \cite{torkhani2012} for meshes, and the point-to-point,  point-to-plane and plane-to plane distances \cite{Tian2017a, Alexiou2018d} and PC-MSDM \cite{Meynet2019a} for point clouds.
The following works pioneered the development of metrics for 3D content with color attributes, some of which were designed for meshes \cite{Tian2008, Guo2016,Nehme2021TVCG} and the majority for point clouds such as PCQM \cite{Meynet2020a},  Hist\_Y \cite{Viola2020a},  GraphSIM \cite{Yang2020a, ZhangY2021a},  Point-to-distribution \cite{Javaheri2020b}.

With the rise of machine learning, a new category of quality metrics has emerged. These metrics are based on a purely data-driven approach, and do not rely on any explicit model.  They learn/optimize the weights of geometric and color descriptors using mainly regression \cite{Chetouani2018, Nehme2021TVCG, Yildiz2020}. 
More recently, deep learning approaches are gaining in popularity. They allow, among other benefits, the emergence of no-reference methods \cite{Nouri2017}.
Convolutional Neural Networks (CNNs) were investigated and adjusted to assess the quality of both meshes and point clouds. In \cite{Abouelaziz2017}, the CNN was fed with perceptual hand-crafted geometric features extracted from the 3D mesh and presented as 2D patches. An end-to-end sparse CNN was designed to develop a no-reference quality metric for colored point clouds \cite{Liu2021}. 
A recent work computed the perceptual loss for point clouds using an auto-encoder architecture based on convolution layers \cite{Quach2021a}.

The field of quality assessment of 3D content, especially those with color attributes (either in the form of texture maps or vertex/point colors), can still be considered to be in its early stages compared to that of images.  Thus, many image-based approaches have been proposed for the quality assessment of 3D data,  whether meshes \cite{Zhu2010, Caillaud2016} or point clouds \cite{Yang2020b, He2021a,Wu2021a}, using existing well-known Image Quality Metrics (IQMs), such as SSIM (and its derivatives) \cite{Wang2004a}, iCID \cite{Preiss2014}, HDR-VDP2~\cite{Mantiuk2011}, VIF \cite{Sheikh2006a}, etc.
That is, IQMs are applied to projected views (rendered snapshots) of 3D models allowing a complete capture of geometric and color distortions as reflected in the final rendering as well as environmental and lighting conditions. 
Lately, several authors have exploited CNNs to assess the quality of 3D content using image-based approaches. For instance, PQA-Net \cite{Liu2021b} is a deep neural network for no-reference quality evaluation of point clouds that extracts features from multiple views using CNNs. The features are then fused and fed to a distortion identifier and a quality predictor.
Another network, proposed in \cite{Tao2021a} for colored point clouds, uses a feature extractor composed of sequential CNNs to extract multiscale features from geometry and color patches separately.  The final quality score is then obtained as a weighted average across all patches.
For meshes,  a recent metric was devised by extracting feature vectors from 3 different CNN models and combining them \cite{Abouelaziz2020}. It uses a patch-selection strategy based on mesh saliency to give more importance to perceptual relevant/attractive regions. 
The existing works considered geometry-only meshes (without color/texture attributes).  In this work, we propose a learning-based quality metric for textured meshes.

Over the last years, convolutional neural networks have successfully rivaled traditional image quality metrics \cite{Gao2017,  zhang2018, Bosse2018, Talebi2018}. Readers can refer to \cite{Tariq2020} for a comprehensive study determining why deep features are good predictors of image quality. 
For 3D content,  it is still difficult/challenging to develop quality metrics based on deep learning, mainly due to the lack of large and rich datasets of 3D objects, especially those with color attributes, as mentioned previously in subsection \ref{subsec:RelatedWork_subj}. 
Given our large-scale dataset of 3000 textured mesh, we propose a learning-based metric, called Graphics-LPIPS, for assessing the quality of rendered 3D graphics. The metric is an extension of the LPIPS metric \cite{zhang2018} originally designed for images and perceptual similarity tasks, which we adapted for 3D graphics and quality assessment tasks.  Our metric employs pre-trained CNN with learning linear weights on top that we fed with patches of rendered images of  3D models.  Our metric provides a good stability and excellent results in terms of correlation and classification ability on two different datasets.



\section{Dataset generation} \label{sec:TexturedDBGeneration}
We produced a large-scale textured meshes quality assessment dataset composed of 343750 distorted meshes derived from 55 source models each associated with 6250 distorted versions. 
Distortions represent combination of level of details simplification, and texture and geometry compression.
The dataset covers a wide range of geometric, color and semantic characteristics. Indeed, each source model has been carefully selected and characterized as will be shown in this section.

\subsection{3D source model selection}
\begin{figure}[htbp]
	\centering
	\includegraphics[width=\linewidth]{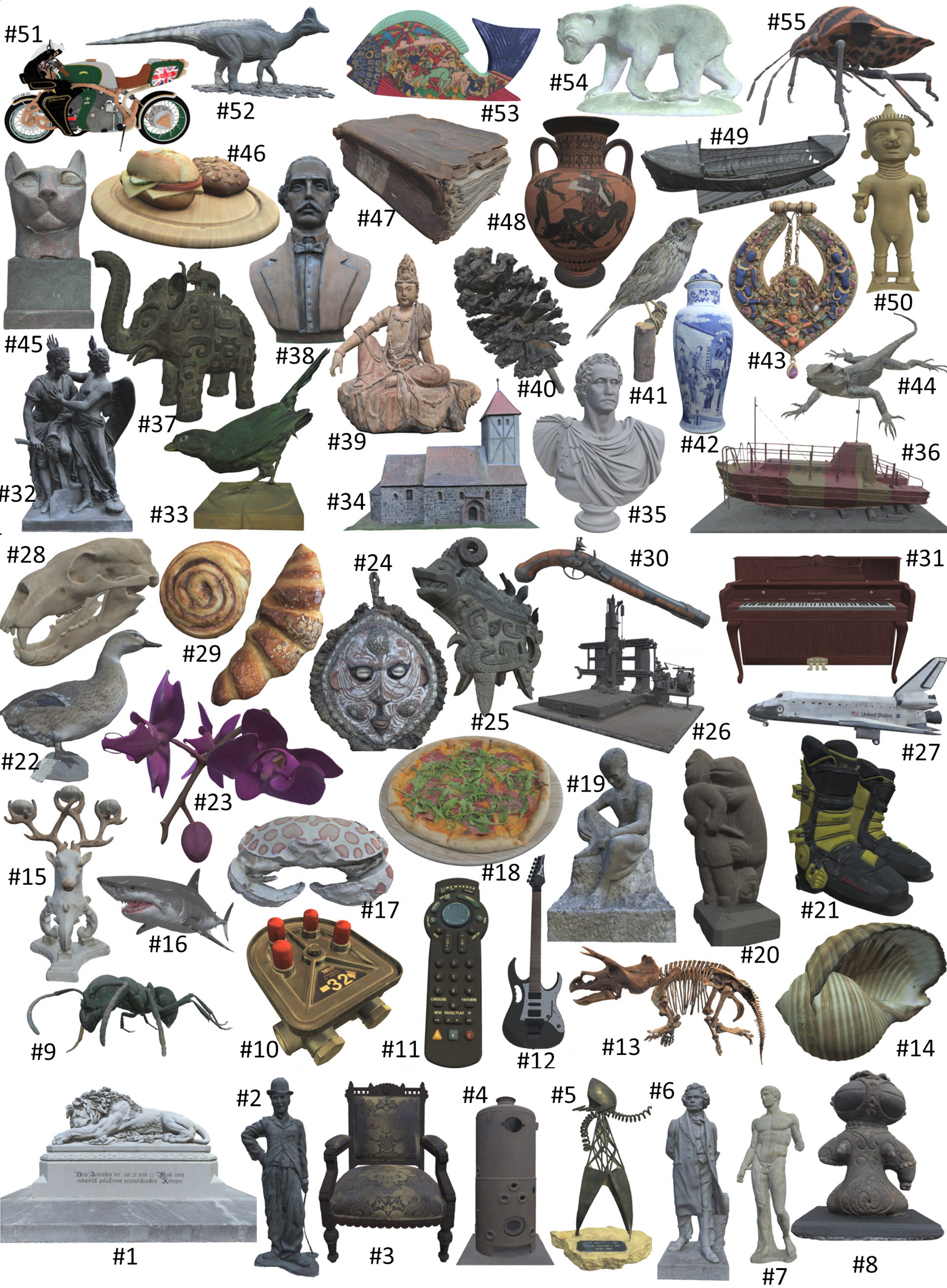}
	\caption{The 3D graphic source models constituting our database.}
	\label{fig:SourceModels}
\end{figure}

We collected 55 textured 3D models from SketchFab\footnote{https://sketchfab.com/features/free-3d-models}. 
The selection was done manually and carefully to get high quality textured meshes with creative commons licenses. Table \ref{tab:sourcemodels} lists the models, their number of vertices and semantic category, while Figure \ref{fig:SourceModels} illustrates them.

Some models were cleaned up to repair topological and geometrical defects (zero-area triangles, non-manifold geometry, holes, etc.). In addition, all models were converted to a unique format: the meshes are provided as OBJ (+ the material file), and the textures as JPEG images of 2K resolution (normalized texture size: $2048 \times 2048$).  The textures encode surface colors (i.e. diffuse map); other information such as surface normals, roughness, and ambient occlusion are ignored. For models with multiple texture images, these were baked into one single image. \modified{More details about the data preparation can be found in supplemental material.}

\begin{table}[htbp]
  \begin{center}
    \caption{List of source models, their number of vertices and semantic category.}
    \label{tab:sourcemodels}
		\resizebox{\linewidth}{!}{
    \begin{tabular}{c | c c || c | c c }
	  \hline
		 \textbf{\makecell{Model\\ID}} & \textbf{\#Vertices} & \textbf{\makecell{Semantic\\category}} &  \textbf{\makecell{Model\\ID}} & \textbf{\#Vertices} & \textbf{\makecell{Semantic\\category}} \\
		\hline
		\textbf{\#1} & 357364 & Animal statue& \textbf{\#29} &  119038 & Food \\
		\textbf{\#2} & 123189 & Human statue & \textbf{\#30} &  114145 & Decoration\\
		\textbf{\#3} & 395490 & Furniture & \textbf{\#31} & 16803 &  Musical instrument \\
		\textbf{\#4} & 99984 & Machine & \textbf{\#32} &  358684  & Sculpture  \\
		\textbf{\#5} & 198683 & Decoration & \textbf{\#33} & 155931 &   Animal statue  \\		
	 \textbf{\#6} & 258490 & Human statue &  \textbf{\#34} & 486850  & Building  \\
	 \textbf{\#7} & 483746 & Human statue &\textbf{\#35} &  150006 &   Bust \\
	 \textbf{\#8} & 250723 & Sculpture & \textbf{\#36} & 249439 & Mean of transport   \\
	\textbf{\#9} & 189633 & Animal & \textbf{\#37} &  130016 &  Sculpture  \\
	\textbf{\#10} & 109929 & Machine &\textbf{\#38} &  304435 &  Bust \\
	\textbf{\#11} & 134472 & Electronic device &  \textbf{\#39} & 100890 & Sculpture  \\
	\textbf{\#12} & 17560 & Musical instrument & \textbf{\#40} & 124936 & Plant \\
	\textbf{\#13} & 75950 & Animal skeleton & \textbf{\#41} & 74819 &  Animal\\	
	\textbf{\#14} & 209609 & Animal & \textbf{\#42} & 150498 & Decoration \\
	\textbf{\#15} & 77924 & Animal statue & \textbf{\#43} & 62989 &  Decoration\\
	\textbf{\#16} & 669346 & Animal & \textbf{\#44} & 36204 & Animal \\
	\textbf{\#17} & 393652 & Animal &  \textbf{\#45} & 650778 & Sculpture\\
	\textbf{\#18} & 27611 & Food & \textbf{\#46} & 4999 & Food \\
	\textbf{\#19} & 308944 & Sculpture & \textbf{\#47} & 110819 & Book  \\
	\textbf{\#20} & 185416 & Sculpture & \textbf{\#48} & 56902 & Decoration  \\
	\textbf{\#21} & 149906 & Apparel & \textbf{\#49} & 92223 & Mean of transport \\
	\textbf{\#22} & 74662 & Animal &\textbf{\#50} &260670& Decoration\\
	\textbf{\#23} & 20144 & Plant & \textbf{\#51} &  60710& Mean of transport \\
	\textbf{\#24} & 297988 & Decoration & \textbf{\#52} & 635206 & Animal statue\\		
	\textbf{\#25} & 151002  & Sculpture & \textbf{\#53} & 150000 &  Decoration \\
	\textbf{\#26} & 98763 &  Machine & \textbf{\#54} &  299976&  Animal \\
    \textbf{\#27} & 77027 &  Mean of transport &  \textbf{\#55} & 125002 & Animal \\
    \textbf{\#28} &  306933 & Animal skeleton & \textbf{} &  & \\
	\hline
    \end{tabular}
    	}
  \end{center}
\end{table}
		
\subsection{Content characterization}\label{sec:contentCharac}
Our goal is to create a high diversity/generality dataset and avoid biases related to the selection of source models. To this end, we proposed an approach to quantitatively characterize the geometric, color and semantic complexity of 3D models.

The characterization of 3D models is not as straightforward as it seems due to the multimodal nature of these data (geometry, color/texture and material information).
In the field of images and videos, the content characterization and classification is usually based on the Spatial perceptual Information (SI) \cite{ITU-TP.910, Yu2013, Engelke2009}. It is an indicator of edge energy, i.e. it emphasizes regions of high spatial frequency that correspond to edges: an image $I$ is converted to gray-scale, then filtered with horizontal and vertical Sobel kernels. The standard deviation ($std$) over the pixels of the Sobel-filtered image is finally computed (Eq. \ref{Eq:SI}).
\begin{equation}
	SI = std_{space}[Sobel(I)]
	\label{Eq:SI}
\end{equation}

Since there is no standard on how to characterize 3D content and inspired by the SI of images, we proposed two new measures to characterize the color and the geometry of textured 3D models. We also introduced a third measure, based on the visual attention complexity \cite{Abid2020} (presented later in this subsection), to characterize the models regarding their semantic aspect.
\modified{Our measures are based on rendered images, which means they are sensitive to the chosen viewpoint. However, for most objects, our human judgment will be mainly focused on the most informative viewpoint i.e. the one containing the most color, geometry, and semantic information. One way to reflect this tendency would be to compute our measures on a variety of viewpoints and take their maximum value. For the sake of simplicity, and because we will further use this main viewpoint to train our network, we manually chose the main viewpoint for each model (shown in Figure \ref{fig:SourceModels}) and compute our measures on renderings from this viewpoint. We conducted experiments (detailed in the supplemental material) that show that this manual viewpoint selection leads to very similar results to max-pooling measures taken from 4 viewpoints around the objects. The proposed characterization strategy can thus be applied in both cases (automatic viewpoint sampling + max-pooling or manual viewpoint selection) with similar results. }\\
\begin{figure}[h!tb]
	\centering
	\includegraphics[width=\linewidth]{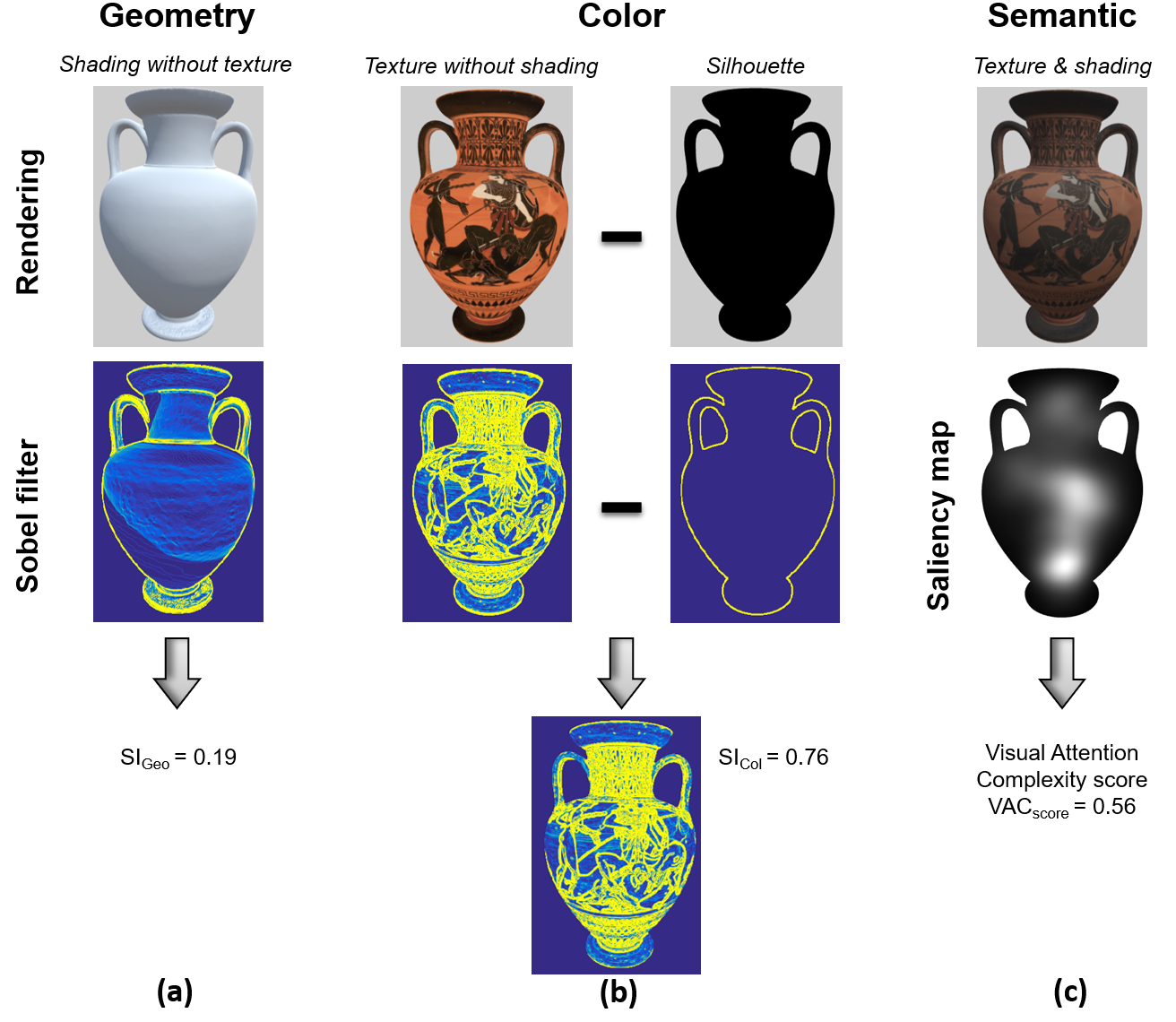}
	\caption{Characterization of the geometry, color, and semantic of textured 3D models.}
	\label{fig:ContentCharac}
\end{figure}
\begin{figure*}[b!]
  \hspace{-5em}
  \begin{subfigure}[h!tb]{0.49\linewidth}
		\centering
		\captionsetup{skip=0pt}
    \includegraphics[width=1.2\linewidth]{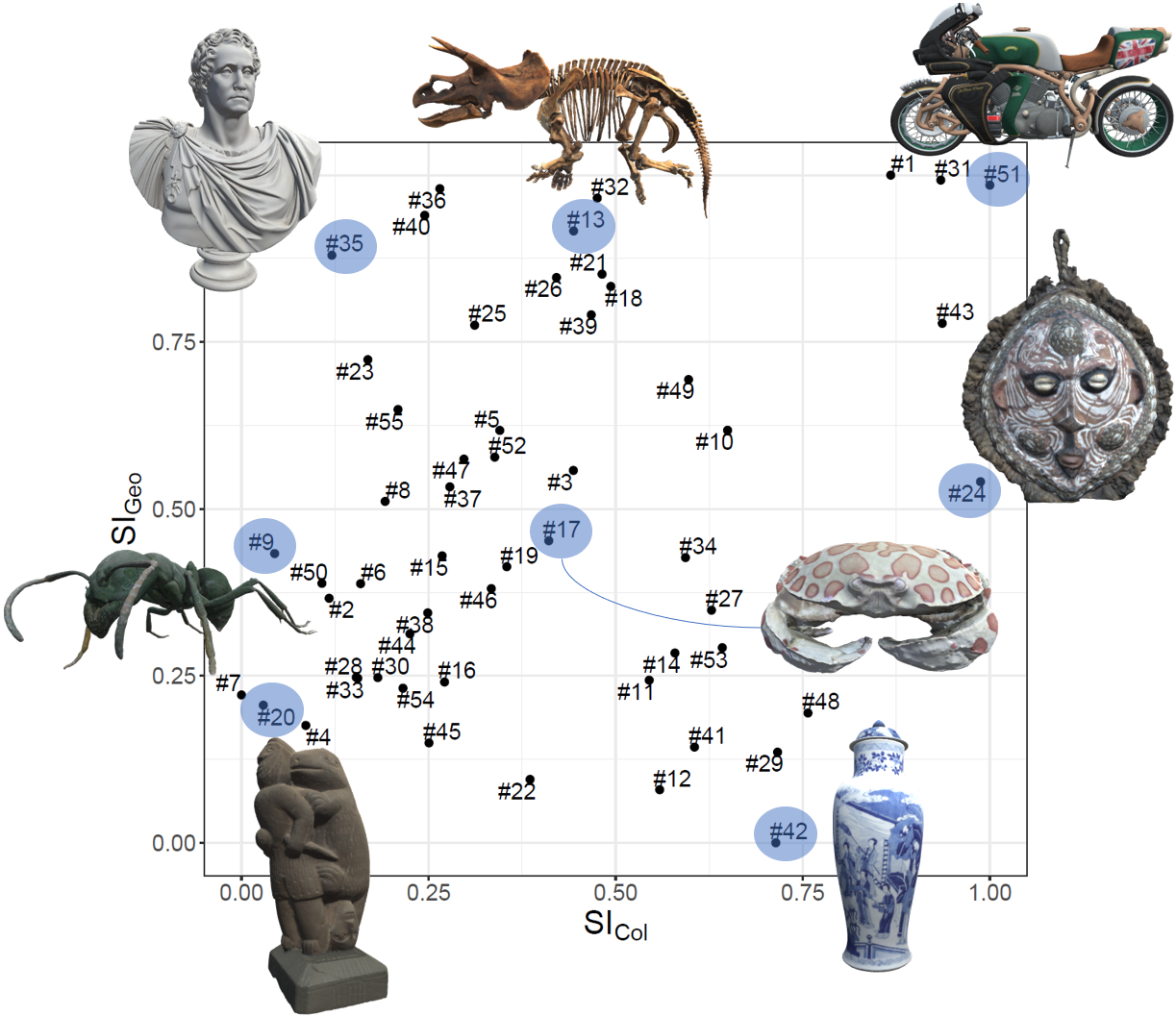}
    \caption{}
  \end{subfigure}\hfill
  \begin{subfigure}[h!tb]{0.49\linewidth}
		\centering
		\captionsetup{skip=0pt}
    \includegraphics[width=1.2\linewidth]{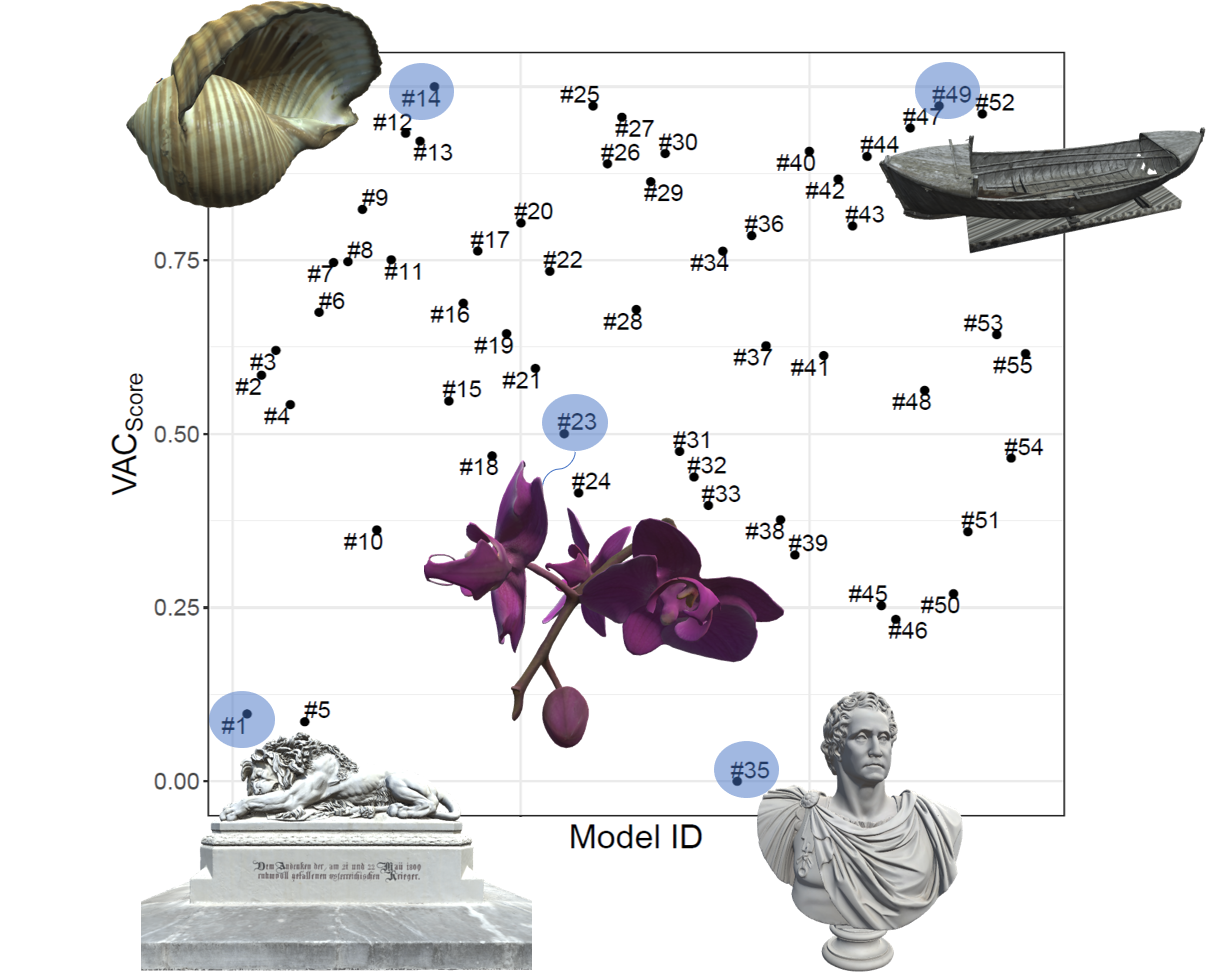}
    \caption{}
  \end{subfigure}
 \caption{(a) Geometry and color spatial information and (b) visual attention complexity for the selected source models.}
 \label{fig:ContentCharac_Results}
\end{figure*}

\textbf{A. Geometric characterization}
To enhance the geometry of the model, we compute a rendering that only takes into account the shading of the model without any of its texture information. Then, we calculate the SI of the rendered snapshot. We thus obtain $SI_{Geo}$ an objective indicator/measure that assesses the geometric complexity of the 3D model.
See Figure \ref{fig:ContentCharac}.a.\\

\textbf{B. Color characterization}
To assess the textural/color characteristics, we consider this time a rendering without any shading; just the colors from the texture are displayed. We apply the Sobel filter on the rendered snapshot and remove the silhouette detected by the filter because it rather reflects geometric information than color information, and is already taken into account by $SI_{Geo}$. The color characterization is thus defined by the $SI_{Col}$ measure computed on the obtained filtered image.
See Figure \ref{fig:ContentCharac}.b.\\

\textbf{C. Semantic characterization}
Semantic information cannot be characterized by the SI. We therefore employed the Visual Attention Complexity (VAC) measure, recently proposed in \cite{Abid2020}, which is perfectly adapted for this task. 
The VAC measure consists in evaluating the dispersion of salient regions of the rendered 3D model. Rendered images of 3D models associated with low VAC scores indicate that there are highly salient regions that attract human visual attention (focused gaze behavior), while images with high VAC scores indicate that the saliency is diffused and not focused on one region (overall gazing behavior).\\
The VAC is computed as follows and illustrated in Figure \ref{fig:ContentCharac}.c: 
First, we render the 3D model with shading and texture attributes under its main viewpoint. Once the snapshot of the final rendered object is generated, we compute the saliency map (which represents the probability of gazing at each pixel) using the ``Salicon" computational model as recommended in \cite{Abid2020}. Finally, we compute a conditional entropy on the saliency map. 
Thus, we obtain the $VAC_{score}$ which characterizes the visual attention complexity of the 3D model and is closely linked to a semantic value.\\

We applied these measures on our 55 selected models. The measures were computed on snapshots of the models having the same resolution\remove{, and normalized between 0 and 1}. 
The obtained \modified{measures, normalized between 0 and 1,} are shown in Figure \ref{fig:ContentCharac_Results}. 
As it can be seen, the source models of our dataset have various geometric, color and semantic characteristics. Figure \ref{fig:ContentCharac_Results}.a shows a good distribution on the $SI_{Geo} / SI_{Col}$ plane: models located at the right-bottom corner of the plane, such as the model \#42, present a rich texture with very low geometric complexity. On the contrary, the model \#35 in the top-left corner of the plot is monochrome but has sharp edges and many small geometric details to depict its hair and face.\\
Figure \ref{fig:ContentCharac_Results}.b shows that our models cover different semantic aspects. For instance, model \#1 and \#35 have a low $VAC_{score}$ indicating that the visual attention of the participants will probably be focused on the salient regions of these models that are the epigraph (the writing more generally) and the face, respectively. On the other hand, there are no particularly salient regions on models \#14 and \#49. These models exhibit low visual attention complexity.\\

The set of measures we proposed reveals the characteristics of 3D models in terms of geometry, color and semantics. These measures are extremely fast to compute and work consistently for both coarse and dense 3D data (regardless of the 3D data representation and the number of vertices). 

\subsection{Distortions}\label{Dist_TexturedDB}
From the 55 source models, we created 343750 distorted versions generated from combinations of 5 real-world distortions. The distortions represent lossy compression and simplification algorithms applied on the geometry, texture mapping and texture image.
\begin{itemize}[leftmargin=*]
\item \textbf{Level of Detail (LoD) simplification}: we applied a surface simplification algorithm based on iterative edge collapse and quadric error metric. This algorithm takes into account both geometry and texture mapping \cite{Garland1998,Meshlab}.
We generated 10 levels of simplification ($LoD_{simpL} \in [L1, L10]$) by uniformly reducing the number of mesh faces so that the mesh of the most degraded level ($LoD_{simpL} = L10$) has around 2000 faces (regardless of the source model). Thus, $\Delta_{simpL} = (NbFaces_0 - NbFaces_{min}) / 10 $ where $\Delta_{simpL}$ is the number of faces removed at each $LoD_{simpL}$ level, $NbFaces_0$ is the number of faces of the source model, and $NbFaces_{min}$ = 2000.

\item \textbf{Quantization}: we uniformly quantized the position of the vertices as well as the coordinates of the texture using Draco\footnote{https://github.com/google/draco}, an open-source library for compressing and decompressing 3D geometric meshes and point clouds. To generate the compressed meshes, we considered 5 levels for each attribute:
\begin{itemize}[leftmargin=*]
\item The quantization bits for the position attribute $qp$ range from 7 to 11 bits ($qp \in [7, 11]$). 
\item The quantization bits for the texture coordinates attribute (a.k.a UV map) $qt$ range from 6 and 10 bits ($qt \in [6, 10]$). The UV map represents the parametrization defined to map the texture image onto the model surface. 
\end{itemize}

\item \textbf{Texture map sub-sampling}: we reduced the size of the texture maps by resampling them using the Lanczos low pass filter. We generated 5 texture sizes ($T_{S}$): $2048 \times 2048$ (the original size), $1440 \times 1440$, $1024 \times 1024$, $712 \times 712$, $512 \times 512$. 

\item \textbf{Texture compression}: we used the JPEG compression which is the most commonly used algorithm for lossy 2D image compression.  We selected 5 texture qualities ($T_{Q}$) obtained by varying the compression level: $90$ (the best quality considered but the least effective compression), $75$, $50$, $25$, $10$ (the lowest texture quality and the highest compression).
\end{itemize}
We note that for each distortion type, the degradation levels were selected to cover a range of high, medium and low quality distorted meshes.
\begin{figure}[htbp]
	\centering
	\includegraphics[width=\linewidth]{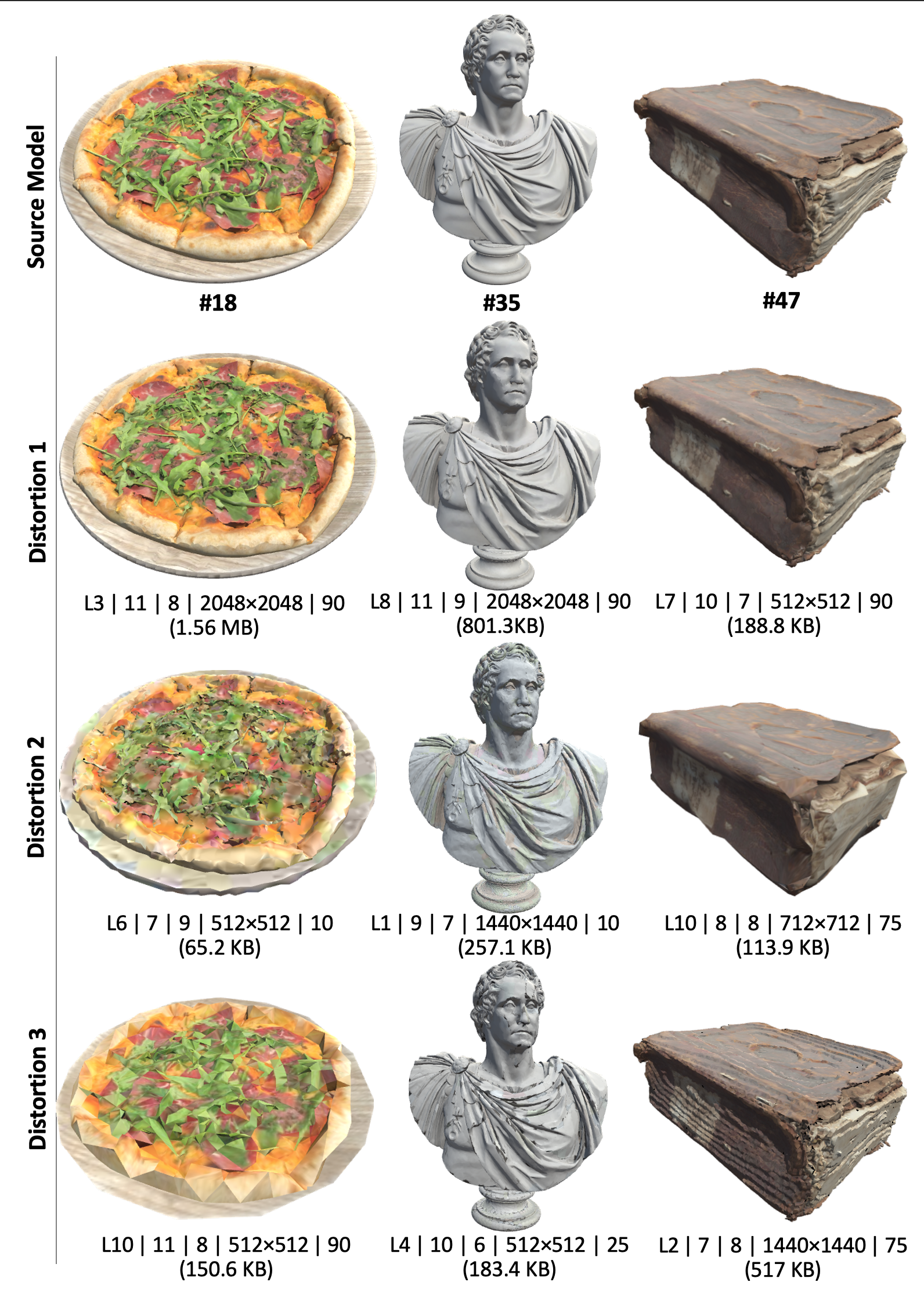}
	\caption{Some examples of distorted models associated with their size (in KB)\modified{, from barely visibles distortions (Distortion 1) up to very annoying ones (Distortion 3)}. Acronyms refer to $LoD_{simpL} \;| \;qp \;| \;qt \;| \;T_{S} \;| \;T_{Q}$.}
	\label{fig:DistortionExamplesTexteuredDB}
\end{figure}

By combining/mixing all geometry and texture distortions, we obtained $10 \;LoD_{simpL} \times 5\;qp \times 5\;qt \times 5\;T_{S} \times 5\;T_{Q} = 6250$ distortions per model, a.k.a. Hypothetical Reference Circuits (HRCs) according to \cite{VQEG2010}. HRCs denote the processing operations applied to the source models to obtain the distorted versions.
Each HRC is associated with a size (in Bytes), resulting from the compression of the source model with the corresponding distortion parameters (using JPEG for the texture and Draco encoder for the connectivity, geometry and UV maps). Thus, the size of a stimulus (in Bytes) is equal to the sum of the size of its compressed texture and its compressed 3D model.

The distortions were applied systematically with the same levels to all models. Our dataset thus contains 343 750 distorted stimuli (55 source models $\times$ 6250 HRCs) that span a great diversity in terms of visual content and quality.
Figure \ref{fig:DistortionExamplesTexteuredDB} shows some visual examples of distorted stimuli along with their distortion parameters. More examples are provided in the supplementary material.

\section{Subjective experiment}\label{sec:TextDB_Exp}
We conducted a large-scale subjective quality assessment experiment in CrowdSourcing (CS), wherein 4513 participants were involved to rate the perceived quality of a generalized and challenging subset of 3000 stimuli carefully selected from the dataset presented above. Over 148k quality scores were collected.
This section describes our extensive online subjective study.

\subsection{Test stimuli selection}\label{subsec:StimuliSelection}
Our dataset contains more than 343k stimuli. Participants cannot be asked to rate the quality of such a large amount of data. We therefore had to select a subset of stimuli to rate in the subjective experiment. The selection of this subset is a crucial step since we aim to use it later to train a learning-based metric. Thus, we seek to obtain an unbiased, generalized and challenging subset, which leads to several selection criteria:
the selected subset had to contain all the source models, as well as an equal distribution of HRCs (combinations of distortions created). In addition, the subset of stimuli must equitably cover the entire range of quality (from imperceptible to very annoying distortions) to have a balanced representation of the visual quality. Last but not least, we want this subset to be challenging for objective quality metrics.

\begin{figure}[htbp]
  \centering
	\includegraphics[width=0.9\linewidth]{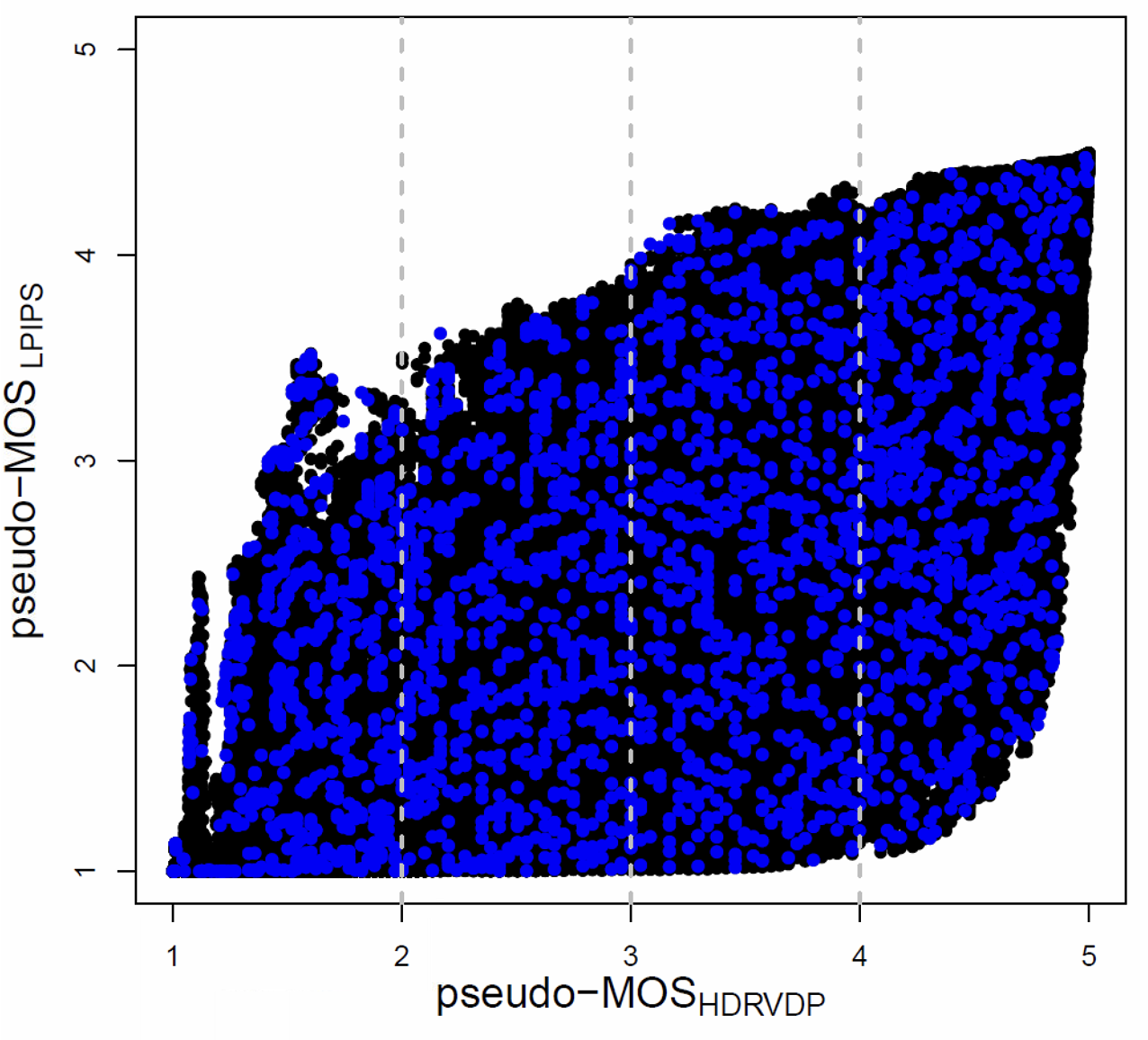}
 \caption{Selection of the test stimuli by constrained sampling of the plane formed by 2 pseudo-MOSs. The black dots refer to the pseudo-MOS values of all stimuli in the dataset, while the blue dots refer to those selected for the subjective experiment.}
  \label{fig:PVSselection}
\end{figure}

We selected 3000 stimuli from 343750 (which represents about 0.9\% of the total dataset).
To do so, we developed the following approach.
\begin{itemize}[leftmargin=*]
\item First, we predicted the Mean Opinion Score (MOS) of all the 343750 stimuli, using existing objective quality metrics that we calibrated on an existing subjectively-rated quality assessment dataset: we used the dataset of meshes with vertex colors \cite{Nehme2021TVCG}. We fitted two logistic regression models (mapping functions) between the MOSs of this dataset and the following two objective quality metrics: (1) HDR-VDP2 \cite{Mantiuk2011} since it provided the best performance among the Image Quality Metrics (IQMs) tested on this dataset \cite{Nehme2021TVCG} and (2) LPIPS (Learned Perceptual Image Patch Similarity) \cite{zhang2018} since it is a commonly used IQM, based on pre-trained CNNs, with many successful applications \cite{Huang2018, Yang2018}. \\
We computed HDR-VDP2 and LPIPS on snapshots of the stimuli in our dataset rendered from their main viewpoints (defined in subsection \ref{sec:contentCharac}), and then predicted their MOS using the obtained regression models.  As in \cite{Liu2021, Wu2020}, we refer to the predicted MOSs as Pseudo-MOSs. Thus,  we obtained 2 pseudo-MOS values per stimulus (pseudo-MOS$_{HDRVDP}$ and pseudo-MOS$_{LPIPS}$).\\
We used two metrics (instead of one) to be able to sample stimuli for which the metrics do not agree on their quality (as shown later in this subsection and in Figure \ref{fig:PVSselection}), resulting in a more challenging subset of stimuli.

\item Second, to ensure a good and equitable coverage of the whole visual quality range and to get a subset of challenging stimuli, we regularly sampled the plane formed by the 2 pseudo-MOSs, shown in Figure \ref{fig:PVSselection}, while respecting 2 constraints: considering HDR-VDP2 as the pivot metric,  we uniformly sampled the area of each quality range. 
For each sampled point, we selected the ``closest" set of stimuli (in terms of distance), and then chose the one that ensures an equal/equitable distribution of (1) all source models (e.g., as many degraded stimuli for model $ID_i$ as for model $ID_j$) and (2) all levels of each distortion(e.g, almost as many stimuli are selected with a $qp = 7$ as those with a $qp = 10$).
\end{itemize}

\subsection{Rendering} \label{sub:rendering}
To adequately assess the visual quality of 3D content, it is important that the object moves so that the observer can see the whole object and not focus on one single viewpoint \cite{Rogowitz2001}. Thus to avoid manual selection of multiple relevant viewpoints for each model, we animated all models in our dataset with a full rotation (360 \textdegree) around their vertical axis. \modified{We rendered the dynamic test stimuli in a neutral room (light gray walls), without shadows and under a directional light coming from the top right of the room}. 
All models were approximately the same size and rendered with a lambertian material\modified{; mipmapping was activated. }

We designed the experiment based on the Double Stimulus Impairment Scale (DSIS) method, in which observers see the source model and the same model impaired side by side and rate the impairment of the second stimulus in relation to the source model using a five-level impairment scale, displayed after the presentation of each pair of stimuli. This method has proven to be the most accurate and stable for evaluating the quality of 3D graphics \cite{Nehme2020TAP}.
Since the experiment is conducted in crowdsourcing, we generated videos of the rendered stimuli in order to limit the participant's interactions with the 3D objects, since we have no full control over the participant's test environment. The only interaction required by the participant is the selection of the score when rating.
The videos were all in $650 \times 550$ resolution (so that the videos of the source and degraded models fit simultaneously on a screen with a minimum resolution of $1920 \times 1080$) with a frame rate of 30 fps and encoded using H.264 encoder (mp4 container) at a bitrate of 5 Mbps to ensure imperceptibility of compression artifacts. Videos are 8 seconds long, which is the time it takes for models to complete the full rotation.

\subsection{Experimental environment}
\begin{figure}[hptb]
	\centering
	\includegraphics[width=\linewidth]{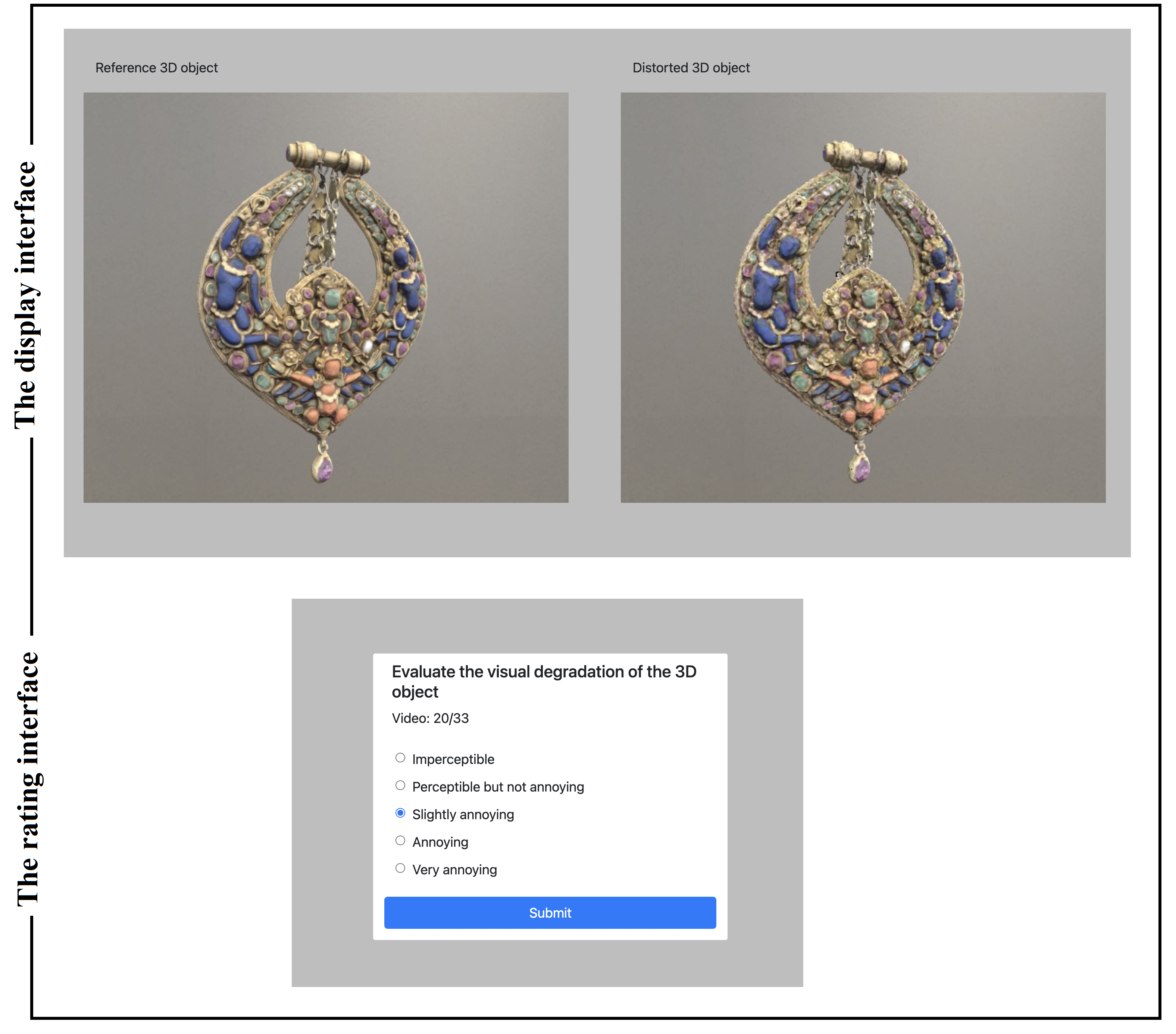}
	\caption{The graphical interface of the subjective experiment platform.}
	\label{fig:ExpEnvironment_TexturedDB}
\end{figure}

To obtain reliable and controlled results, we used our own web platform to present the subjective experiment to participants. This platform has proven its effectiveness in achieving the accuracy of a laboratory test and producing reliable results \cite{Nehme2021MMSP}.  The crowdsourcing service was used only to recruit participants. 
Our platform, illustrated in Figure \ref{fig:ExpEnvironment_TexturedDB}, is suitable for presenting videos according to the DSIS method. 
Only a web browser with an MPEG-4 decoder is required to run the experiment; no other software needs to be installed.
The platform first checks the compatibility of the participant's device:
minimum screen resolution of $1920 \times 1080$, page zoom level, maintain full screen mode throughout the experiment.
The test instructions are then displayed to the participant with a progress bar, at the bottom of this page, showing the status of the loading process of all the video pairs that will be used in the test. This way, the videos of the source and distorted models are played simultaneously during the test, without any latency or unintended interruptions. When the loading is completed a start button appears leading to the test. These steps/windows are illustrated in the supplementary material.
The pairs of videos are displayed in a random order to each participant. Participants cannot replay the videos or provide their score until the videos have been played completely. 
There is no time limit for voting and videos of the stimuli are not shown during that time. At the end of the experiment, participants will receive unique codes allowing them to get their remuneration. 

\subsection{Test sessions \& participants}
\textbf{Creation of test sessions.} 
The number of stimuli and the duration of the experiment should be limited as much as possible in crowdsourcing to keep participants motivated and to avoid unreliable results \cite{Redi2015, Jimenez2018, Reimann2021}.
Therefore, we divided our subset of 3000 test stimuli into 100 playlists. Each participant rates one playlist, i.e., only 30 stimuli. This way, we stay within the duration of the experiment in \cite{Nehme2021MMSP} and within that of a pilot experiment we conducted to evaluate the number of scores needed per stimulus to have the same confidence intervals as for a lab study (see supplementary material).
The test stimuli were fairly distributed among the playlists so that each playlist contains a maximum diversity of source models (a source model is repeated a maximum of 2 times in a playlist). Additionally, each playlist spans the full range of distortions and all playlists have nearly the same pseudo-MOS distribution.

As in \cite{Bovik2015}, we injected 3 Golden Units (GU, a.k.a trapping stimuli) into each playlist to facilitate the detection of unreliable participants later. The golden units (GUs) included (1) a very poor quality stimulus ($GU_{poor}$) ,  (2) a very high quality stimulus ($GU_{high}$) and (3) a stimulus displayed twice ($GU_{rep1}$ and ($GU_{rep2}$) ) to assess the participant's consistency (coherence of his/her scores). Participants who fail to answer correctly the golden units are considered outliers and their scores are discarded.\\
\newline
\textbf{Training.}
The experiment started with training. In order to familiarize participants with the task and the rating scale, we selected 5 stimuli not included in the 3000 stimuli test and all referring to the same source model. Each stimulus represented one level of the five-level scale of the DSIS method.
After displaying each pair of training videos for 8 sec, the rating interface is displayed for 5 sec and the proposed score assigned to this distortion is highlighted. Once the training is completed the actual test began.\\
\newline
\textbf{Duration.}
The test session of our experiment consisted of 33 pairs of videos to rate (1 playlist of 30 stimuli and 3 golden units) and lasted about 10-12 minutes: informed consent $+$ loading videos $+$ instructions $+$ 5 training stimuli $\times$ (8s video length $+$ 5s Rating) $+$ 33 test stimuli $\times$ (8s video length $+ \sim$ 4s Rating).\\
\newline
\textbf{Participants.}
We ran our experiment until each playlist was fully rated by 45 participants. To set the number of participants per playlist, we referred to the CS experiment in \cite{Nehme2021MMSP}, but we also conducted another pilot experiment with 30 stimuli (selected from this dataset) that were rated by 60 participants. We assessed the evolution of the confidence intervals according to the number of participants (see supplementary material).

It took us about 5 days to collect all the data: 148929 quality judgments were collected.
A total of 4513 participants completed the experiment: 2501 males and 2012 females. They were from 67 different countries and aged between 18 and 80. All participants were naive about the purpose of the experiment.
The recruiting process of the participants was conducted using Prolific\footnote{https://www.prolific.co/}, as the results of the CS experiment in \cite{Nehme2021MMSP} highlight the reliability and seriousness of the Prolific participants.\\

\modified{In the remainder of the paper, subjective scores are mapped on a discrete numerical scale from 1 to 5, following the ITU recommendation [ITU-R BT.500-13 2012] as follows: Imperceptible: 5, Perceptible but not annoying: 4, Slightly annoying: 3, Annoying: 2, Very annoying: 1.}
\newline

\textbf{Participants screening.} 
As in \cite{Christian2014}, participants were filtered by combining the following two screening strategies: \\
\noindent(1) the ITU-R BT.500-13 screening procedure \cite{ITU-BT.500-13}, which detected 159 outliers. This procedure is summarized in \cite{Mantiuk2012} as follows: "\textit{The procedure involves counting the number of trials in which the result of the observer lies outside $\pm2\times \textrm{standard deviation}$ range and rejecting those observers for which (i) more than $5\%$ of the trials are outside that range; and (ii) the trials outside that range are evenly distributed so that the absolute difference between the counts of trials exceeding the lower and the upper bound of that range is not more than $30\%$.}"\\
\noindent(2) the golden units (GU) analysis, which revealed 110 outliers distributed as follows:
\begin{itemize}
	\item 24 participants rated the distortion of the very poor quality stimulus as imperceptible or perceptible but not annoying ($s_i^{GU_{poor}} \geq 4$, where $s_i^{GU_{poor}}$ denotes the score assigned by participant $i$ to $GU_{poor}$).
	\item 39 participants rated the very good quality GU as annoying or very annoying ($s_i^{GU_{high}} \leq 2$).
	\item 32 participant gave inconsistent scores to the third GU showed twice ($\left|s_i^{GU_{rep1}} - s_i^{GU_{rep2}}\right|\geq3$).
	\item 7 participants rated $s_i^{GU_{poor}} =3$ \& $s_i^{GU_{high}}= 3$.
	\item 8 participants scored $(s_i^{GU_{high}} =3$ $|$ $s_i^{GU_{poor}} =3)$ \& \\$\left|s_i^{GU_{rep1}} - s_i^{GU_{rep2}}\right| = 2$.
\end{itemize}
Of the participants who failed to evaluate the golden units, 14 were also detected by the ITU-R BT.500-13 screening procedure. As a result, 255 out of 4513 participants were rejected (5.6\%). 
Only the scores of the remaining participants will be used in the following sections. \modified{We compute the Mean Opinion Score (MOS) from these scores by averaging the scores given by different participants on each stimuli.}\\

\begin{figure}[htpb]
  \begin{subfigure}{0.42\linewidth}
    \centering
	\includegraphics[width=\linewidth]{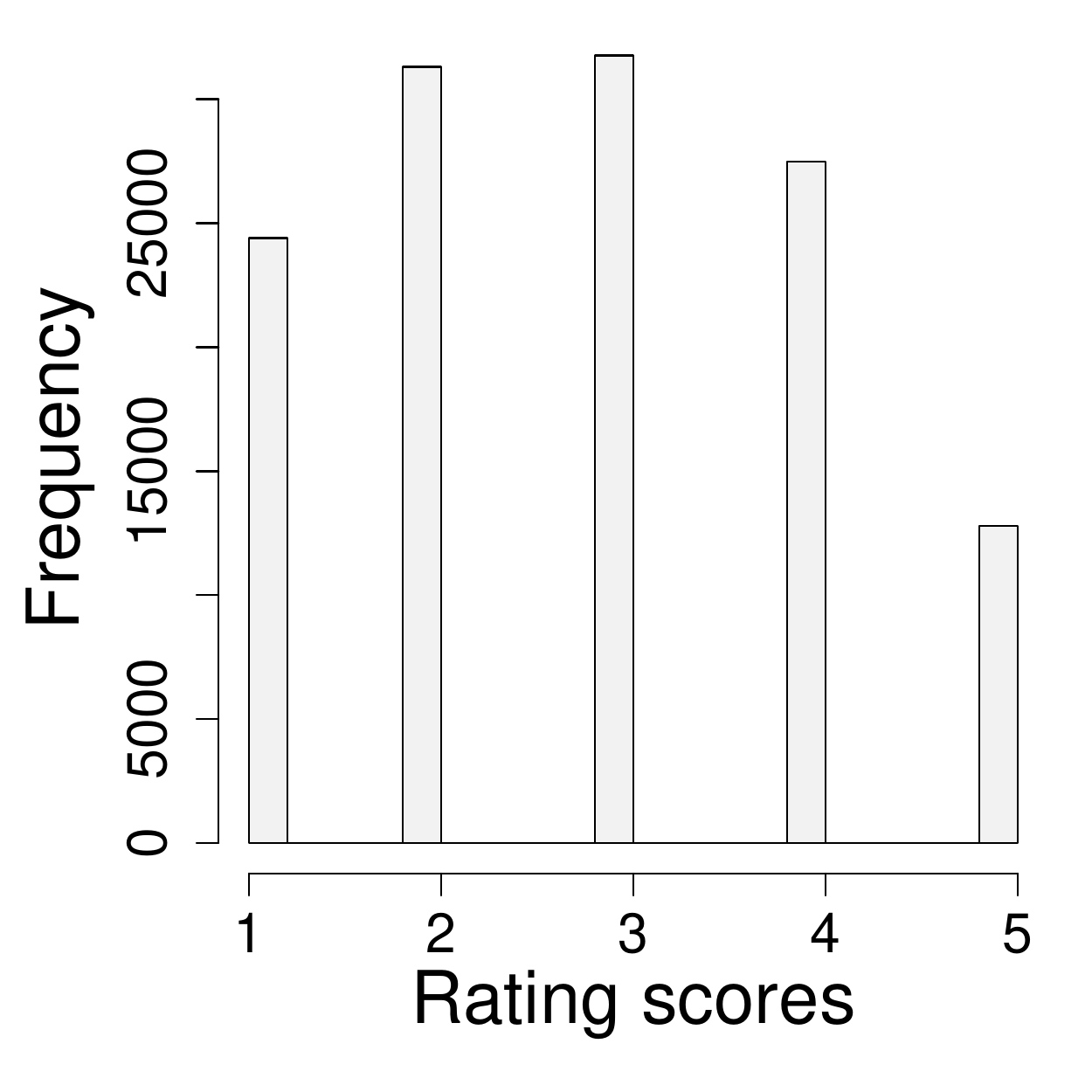}
	\caption{}
\end{subfigure} \hfill
\begin{subfigure}{0.42\linewidth}
    \centering
	\includegraphics[width=\linewidth]{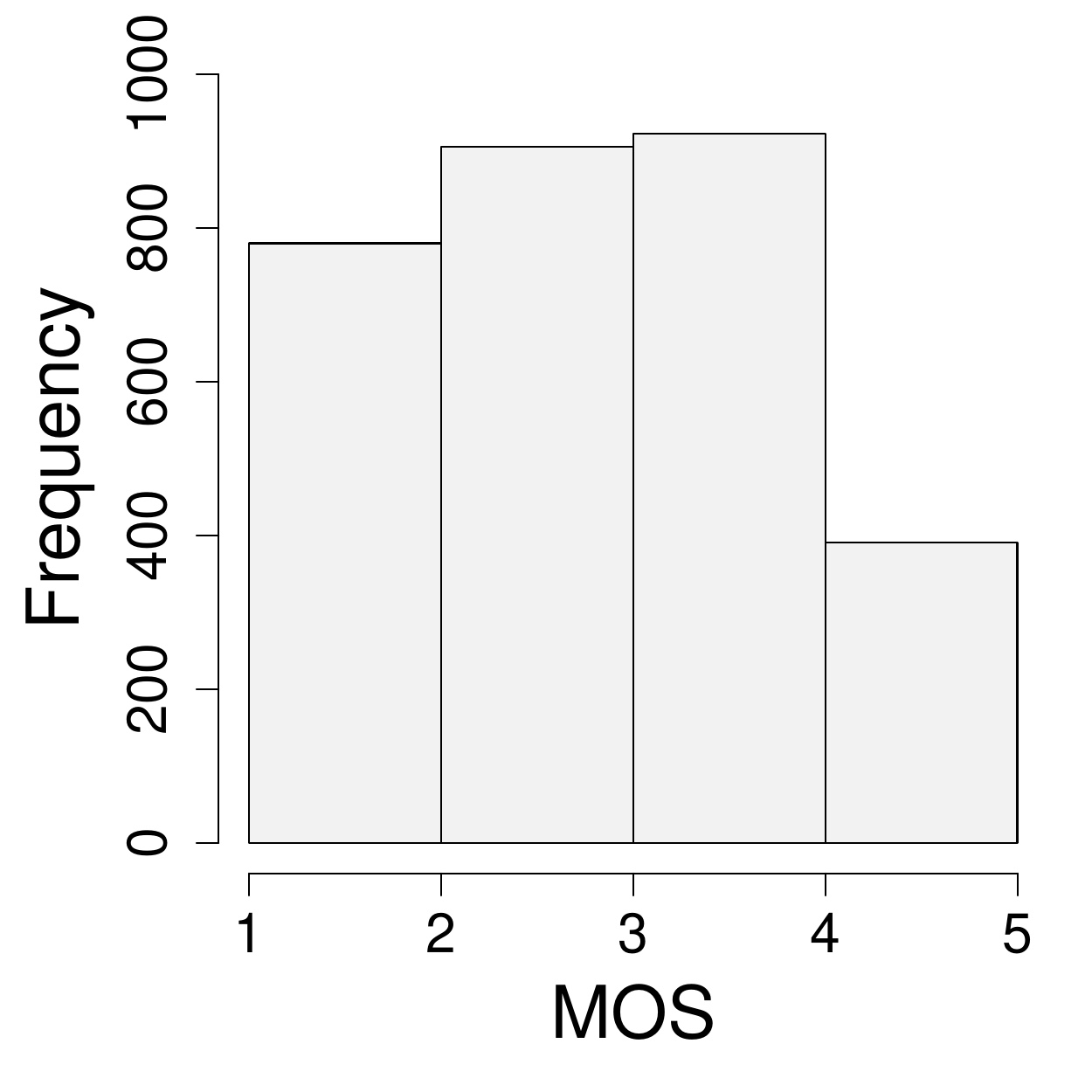}
	\caption{}
\end{subfigure}
\caption{Distribution of (a) raw scores and (b) MOSs for the subset of 3000 stimuli rated in the subjective experiment.}
\label{fig:Playlists_MOS}
\end{figure}

\begin{figure}[htpb]
    \centering
	\includegraphics[width=\linewidth]{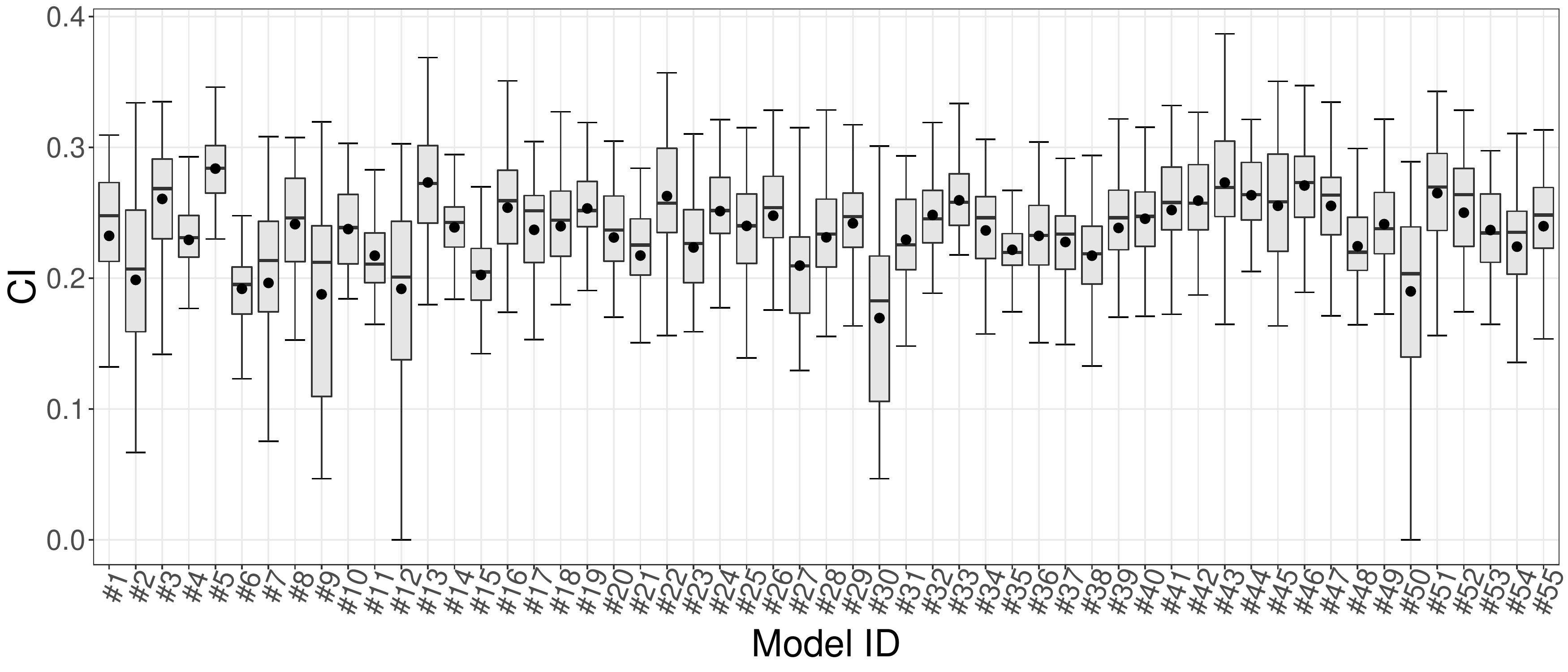}
	\caption{Boxplots of CIs obtained for the source models. Mean values are displayed as circles.}
	\label{fig:Playlists_CI}
\end{figure}

We present in Figure \ref{fig:Playlists_MOS} the distribution of raw scores and MOSs obtained for the subset of 3000 stimuli evaluated by the participants. 
It can be seen that the subjective scores distributed across the whole quality range.
Figure \ref{fig:Playlists_CI} shows the distribution of Confidence Intervals (CIs) of the MOSs for each source model. 
No loose confidence intervals were found \modified{(most confidence intervals are below 0.3)}, demonstrating good agreement between participants' ratings across the stimuli of the different models.


\section{Graphics-LPIPS: A perceptual quality metric for 3D graphics based on CNN}\label{sec:Graphics-LPIPS}
As discussed in the related work section, there is a lack of quality metrics for 3D graphics with color attributes, especially those based on deep learning approaches. Given our large-scale dataset of textured mesh of which 3000 stimuli are associated with subjective scores, we were able to create a deep-learning quality metric for such data. 

We considered the Learned Perceptual Image Patch Similarity (LPIPS) metric \cite{zhang2018} which employs a deep neural network to evaluate the perceptual similarity between 2 image patches, and adapted it to 3D data, then trained it using our dataset. The choice of LPIPS was motivated by its many successful applications \cite{Huang2018, Yang2018}, the simplicity of the approach and the fact that it is based on an in-depth study across different architectures.

\subsection{Description of Graphis-LPIPS}
\begin{figure}[htbp]
	\centering
	\includegraphics[width=\linewidth]{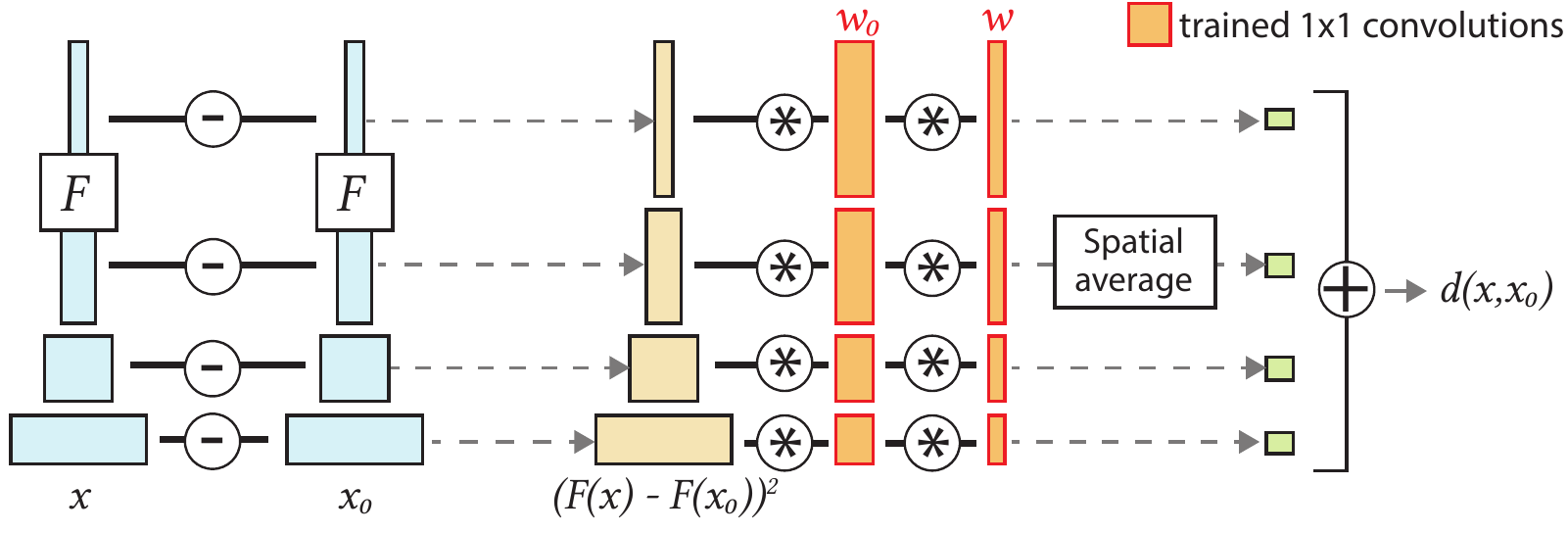}
	\caption{\modified{Graphics-LPIPS architecture: to compute a distance $d(x,x_0)$ between two patches $x$ and $x_0$,  features are first extracted from the layers of the network $F$ and normalized in the channel dimension. The feature difference then goes through two layers of $1\times1$ convolution in order to reach a 1-channel measure. Finally, we average across spatial dimension and sum across all layers.}}
	\label{fig:LpipsOverview}
\end{figure}

\modified{The base principle of LPIPS is to use pre-trained neural networks to extract deep features from two image patches, $x$ (the reference) and $x_0$ (the distorted patch). The two inputs are treated in parallel by two siamese CNNs that share weights and that we denote by $F$. 
The feature difference between the two patches ($F(x) - F(x_0)^2$) can then be mapped to a perceptual difference by going through a $1\times1$ convolution layer that learns the appropriate weights $\omega$ for each channel. Compared to the original LPIPS, we added a $1\times1$ convolution layer with weights $\omega_0$ which allows to further calibrate the model to our perceptual data. Finally, the result is spatially averaged over all pixels from the patch. The obtained score $d(x,x_0)$ represents a perceptual difference between the original patch $x$ and the distorted one. The architecture of our network is shown in Figure \ref{fig:LpipsOverview}. In accordance to the recommendation of the authors, we use the pre-trained AlexNet with fixed weights as feature extractor and only learn the weights $\omega$ and $\omega_0$ of the convolution layers on top.}

\modified{The LPIPS metric was originally trained and tested for perceptual similarity tasks (a.k.a Two Alternative Force Choice 2AFC), in which participants were asked to choose which of two distorted patches ($x_0$, $x_1$) is more similar to the reference $x$. A small network was thus added by the authors at training time to map the obtained perceptual distances to the collected preference score.
Differently from the original LPIPS, our dataset is composed of a MOS scores per image and we want our metric to predict this overall quality score. We thus modify the remaining of LPIPS pipeline in the following ways: (1) we divide our stimuli into patches that are suitable for distance computation, (2) we pool the perceptual distances obtained on all these patches to get an image score that can be compared to the MOS, (3) we devise a training strategy to ensure that the loss can be computed per image (and not only per-patch).}

\modified{Given a distorted image $I$ (along with its collected quality score $MOS_I$) and a reference image $I^r$, we sample corresponding patches $x_i$ and $x_i^r$. We use the previously described pipeline to obtain the perceptual score $d(x_i^r, x_i)$ of each pair of patches.
Similarly to \cite{Bosse2018, kang2014}, we opted for the ``pooling by simple averaging" to get the score of each image, the estimated overall image quality $\hat{Q}$ of $I$ is thus computed as:
 \begin{equation}
 	\hat{Q_I} = \frac{1}{N_p} \sum_{i=1}^{Np} d(x_i^r, x_i)
 	\label{Eq:pooling_simple_averging}
 \end{equation}
 where $N_p$ denotes the number of patches sampled from $I$, $x_i$ refers to a patch from the distorted image $I$, and $x_i^r$ is its corresponding patch on the reference image $I^r$. Note that since our stimuli are renderings of distorted meshes, the perceived visual alterations are expected to be relatively well distributed on the image, without strong local distortions. For this reason, averaging the local scores over the image is a reasonable strategy and gives the best result compared to other pooling functions. Results with other pooling strategies (L2, L3 or max pooling) are reported in the supplementary material.}

\modified{The Mean Square Error (MSE) is used as the minimization criterion. The loss function to train our network is then:
\begin{equation}
	E_I = (\hat{Q_I} - MOS_I)^2
	\label{Eq:G-LPIPS_Loss}
\end{equation}
where $E_I$ is the loss over an image.\\
\newline}

\textbf{Training data.} \label{sec:TrainingLPIPS}
As training data, we use for each stimulus of our dataset a snapshot taken from its main viewpoint to which we assigned the obtained MOS. Thus, we have 3000 annotated images representing our 3000 degraded stimuli. The image size, $650 \times 550$, is the video resolution of the stimuli seen by the participants in the subjective experiment. 
We divided (patchified) the images into small overlapping patches of size $64\times64$ \modified{sampled every 32 pixels}. We removed patches containing less than 65\% stimulus information (i.e., the percentage of background pixels in the patch is greater than 35\%). We got an average of 60 patches per stimulus.

80 \% of the stimuli in the dataset (about 2400) are used for training and 20 \% for testing. 
The dataset is randomly split by source model ensuring that no 3D model is used for both training and testing. As a result 44 source models out of 55 were included in the training while the rest were used for testing. \modified{We used k-fold cross-validation and generated 5 different splits. We report the performances over the 5 folds in the next subsection and chose a representative fold with average performances for all the subsequent evaluations.}

\modified{\textbf{Training.}} \label{sec:TrainingParam}
\remove{We trained the network on randomly sampled patches of the stimulus images. }
As the distances computed for patches of the same image are combined for the optimization \remove{of the weights $\omega$ of the linear layers} of the network (Eq. \ref{Eq:pooling_simple_averging} and Eq. \ref{Eq:G-LPIPS_Loss}), we cannot treat each patch as a separate sample (in other words, the patches of the same image can not be distributed over different batches). Thus, each batch was made to contain $N_I$ images, each represented by $N_p$ patches, resulting in a batch size of $N_I \times N_p$ patches. The backpropagated error is the average loss over the images in a batch (mean($E_I$) computed in Eq. \ref{Eq:G-LPIPS_Loss}).
During training, patches are randomly sampled every epoch to ensure that as many different image patches as possible are used in training.  \modified{At inference time, we use all the patches from the image to make the prediction.} 

\modified{Our final model was trained for 10 epochs (5 epochs at initial learning rate $10^{-4}$ and 5 epochs with linear decay). Each batch contained $N_I = 4$ images (stimuli), each represented by $N_p = 150$ patches. The maximum number of patches for one image in our dataset being 149, all the patches from each image are thus seen at each iteration. For images that are represented by less than 150 patches, we repeat the patches until reaching this number.} 

We refer to the version of LPIPS adapted for 3D Graphics as \textit{Graphics-LPIPS} \modified{and compare its performances to those of the original LPIPS in the following subsection.}

\subsection{Results and Evaluation}\label{sec:TM-LPIPS-results}

\begin{figure}[htbp]
	\centering
	\includegraphics[width=\linewidth, clip, trim=17 17 20 12]{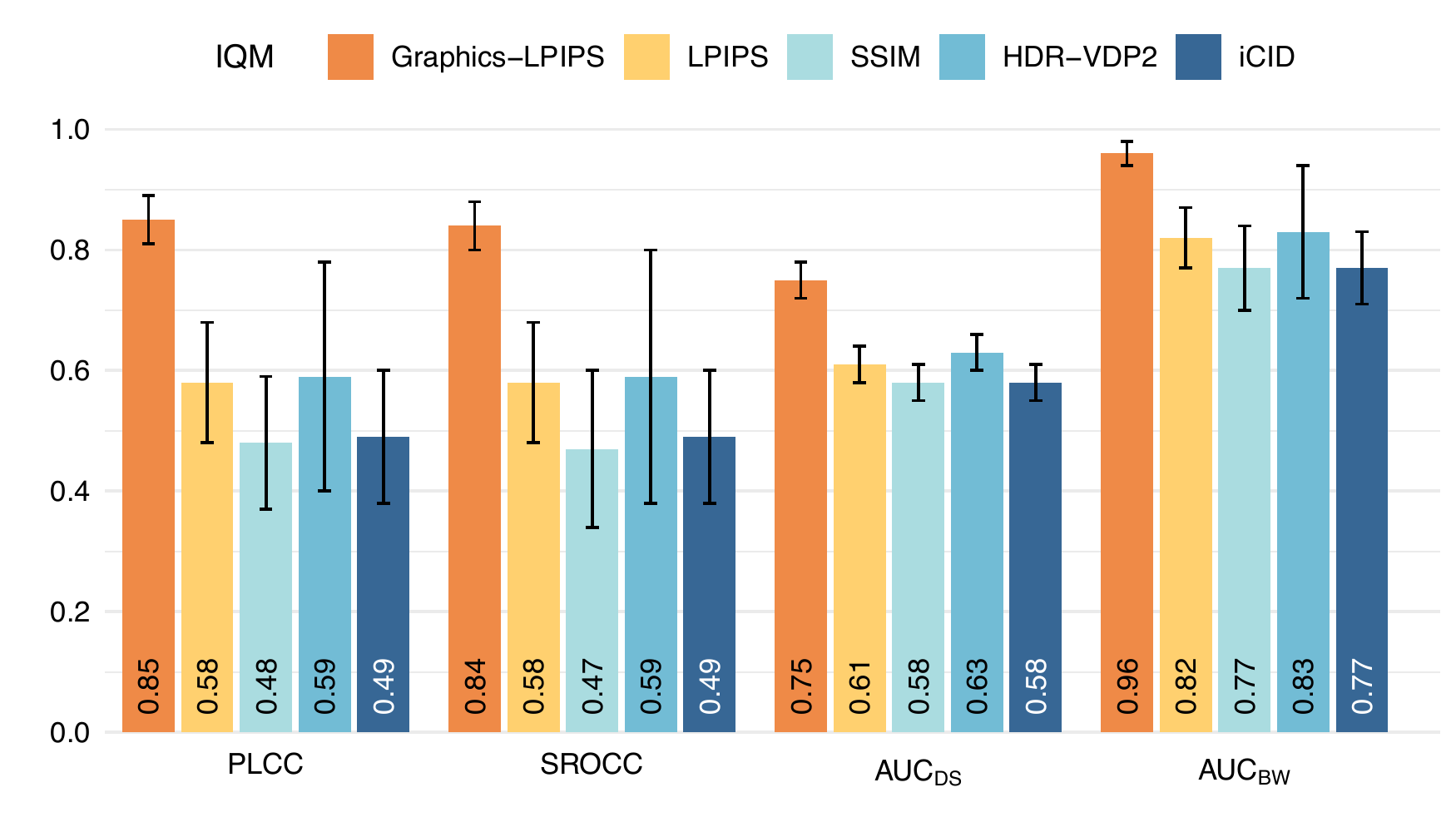}
	\caption{\modified{Performance of our metric (Graphics-LPIPS) compared to state-of-the-art image metrics. The reported numbers are averages over our five folds while the error bars show the standard deviation over the folds.}}
	\label{fig:Perf_testset}
\end{figure}
\begin{figure}[hptb]
  \centering
  \begin{subfigure}[h!tb]{0.5\linewidth}
    \includegraphics[width=\linewidth]{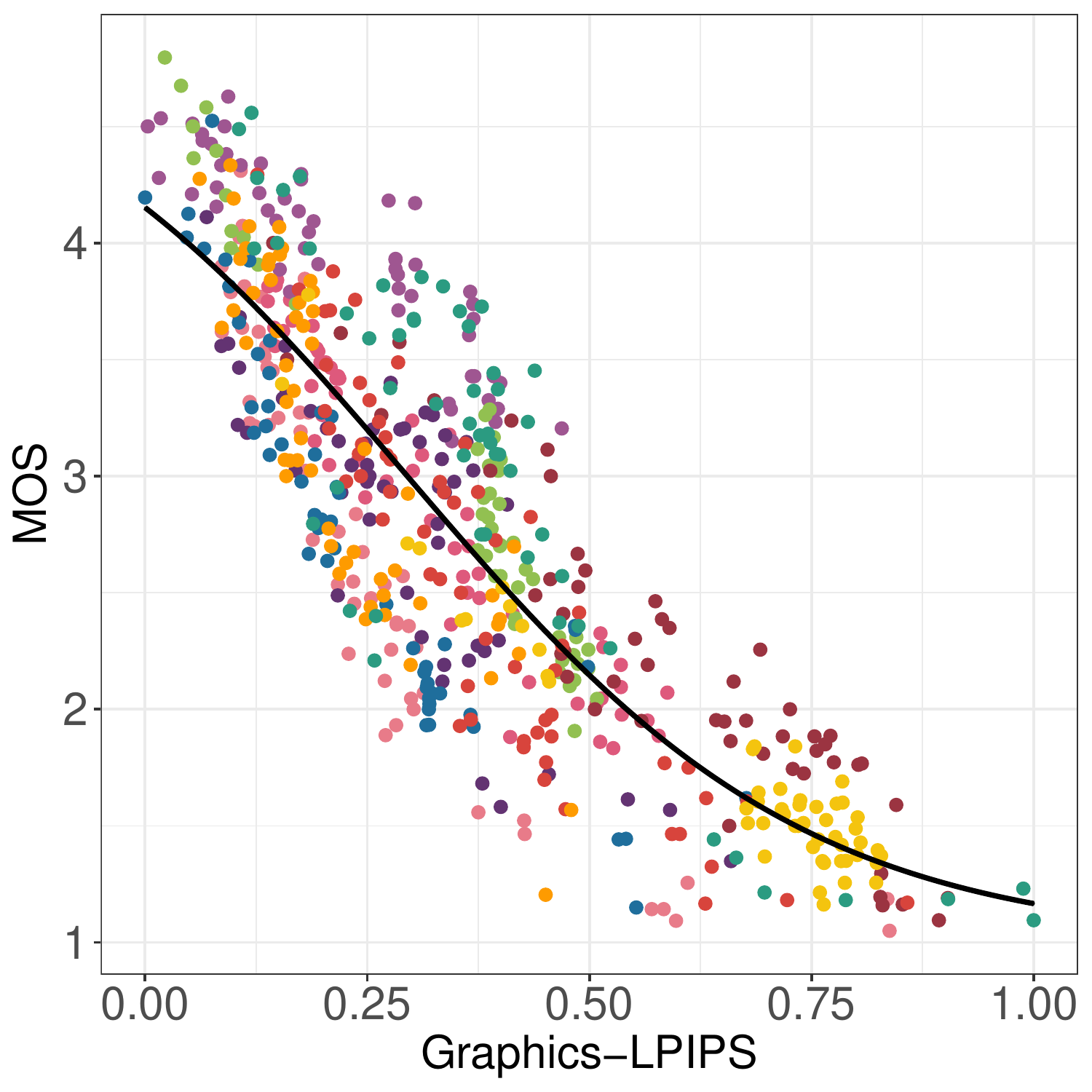}
  \end{subfigure}\hfill
	\begin{subfigure}[h!tb]{0.5\linewidth}
    \includegraphics[width=\linewidth]{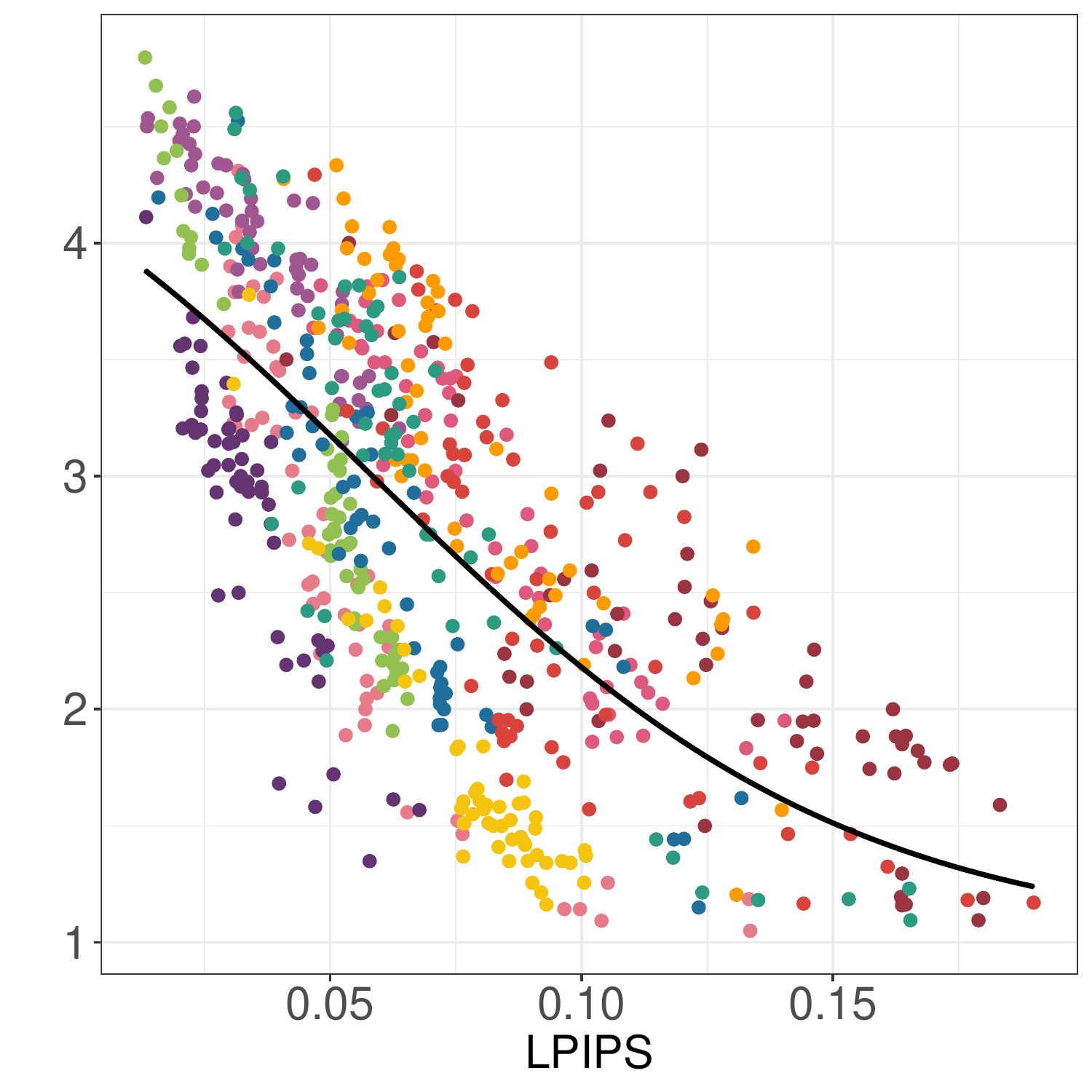}
  \end{subfigure}
   \begin{subfigure}[h!tb]{0.5\linewidth}
    \includegraphics[width=\linewidth]{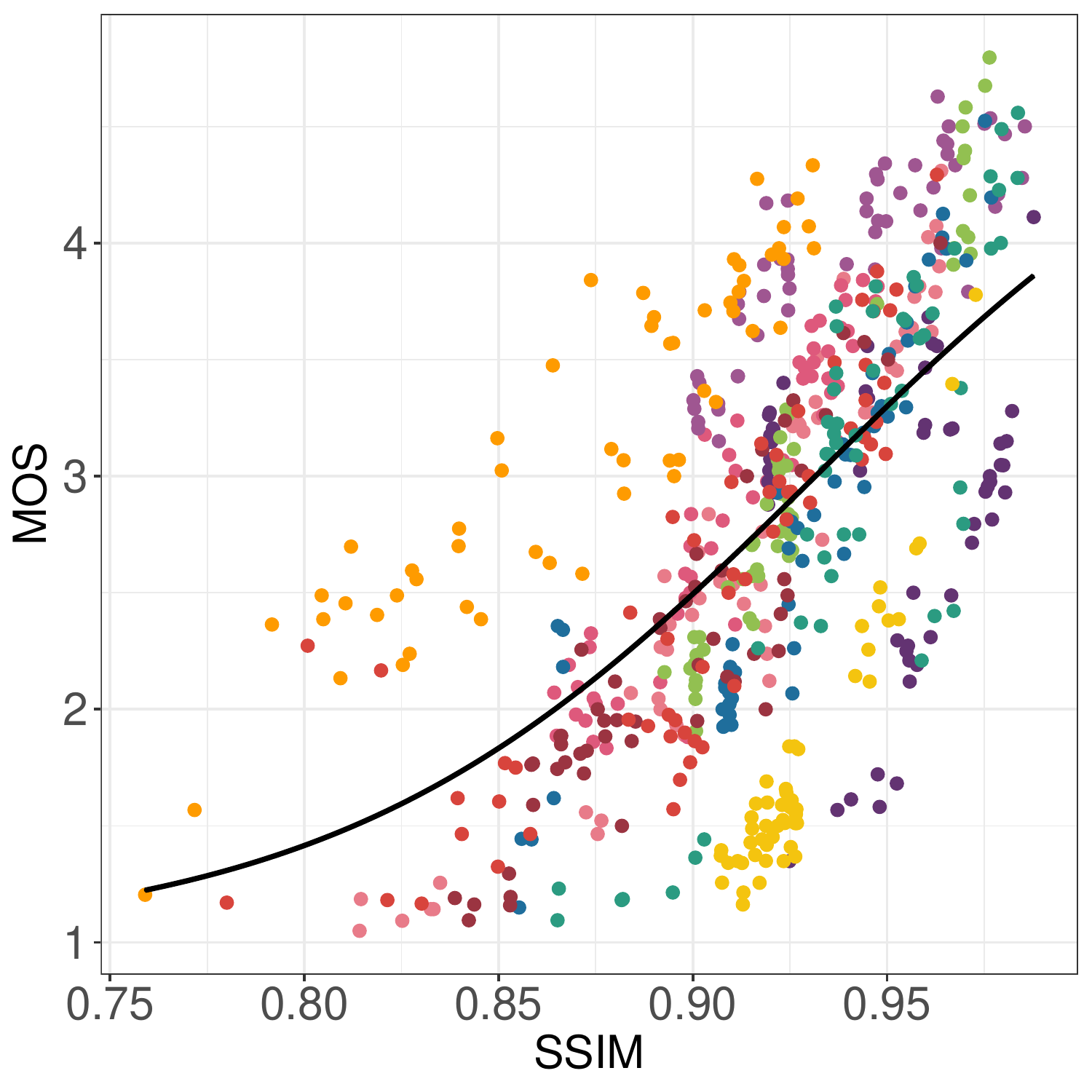}
  \end{subfigure}\hfill
	\begin{subfigure}[h!tb]{0.5\linewidth}
    \includegraphics[width=\linewidth]{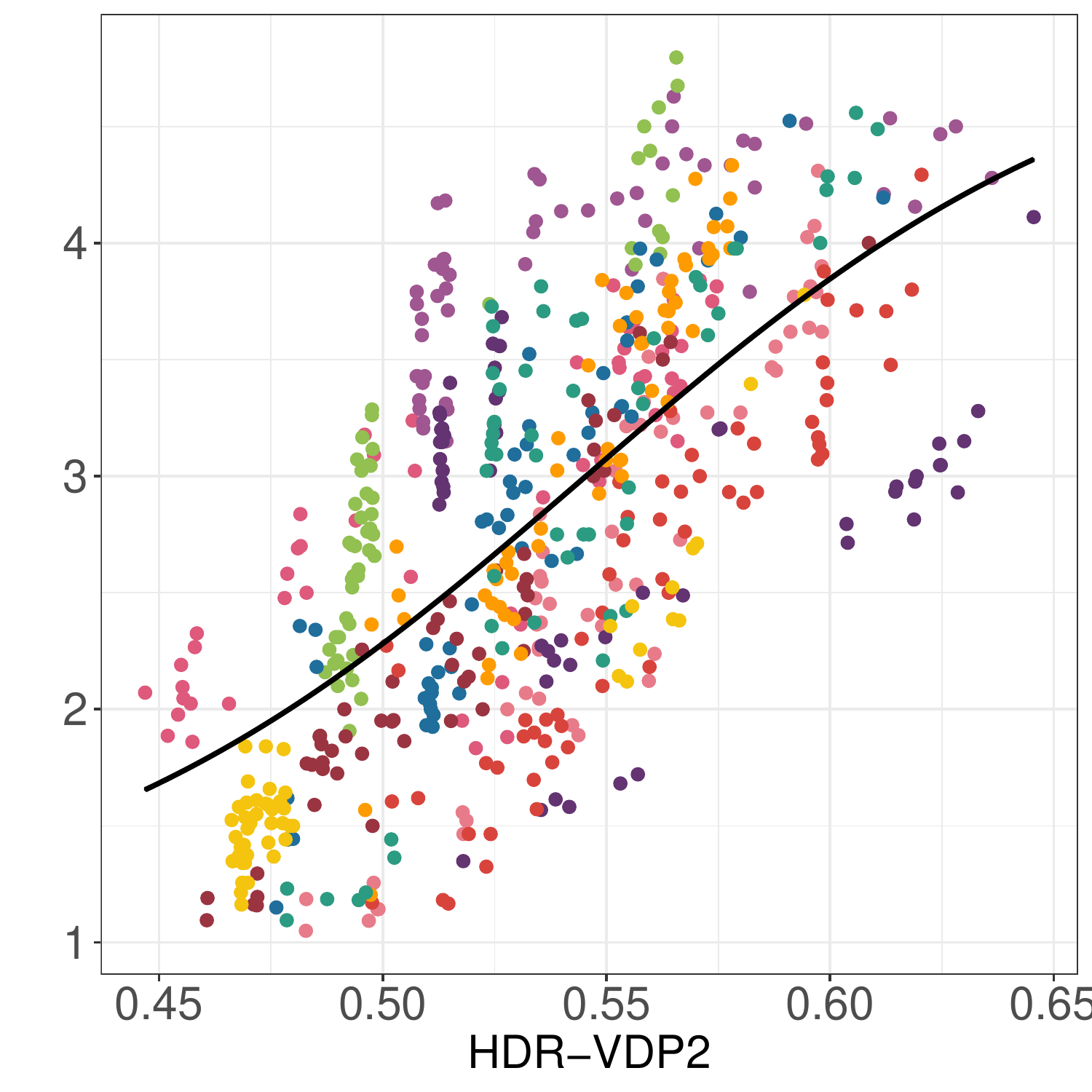}
  \end{subfigure}
  \begin{subfigure}[h!tb]{0.5\linewidth}
    \includegraphics[width=\linewidth]{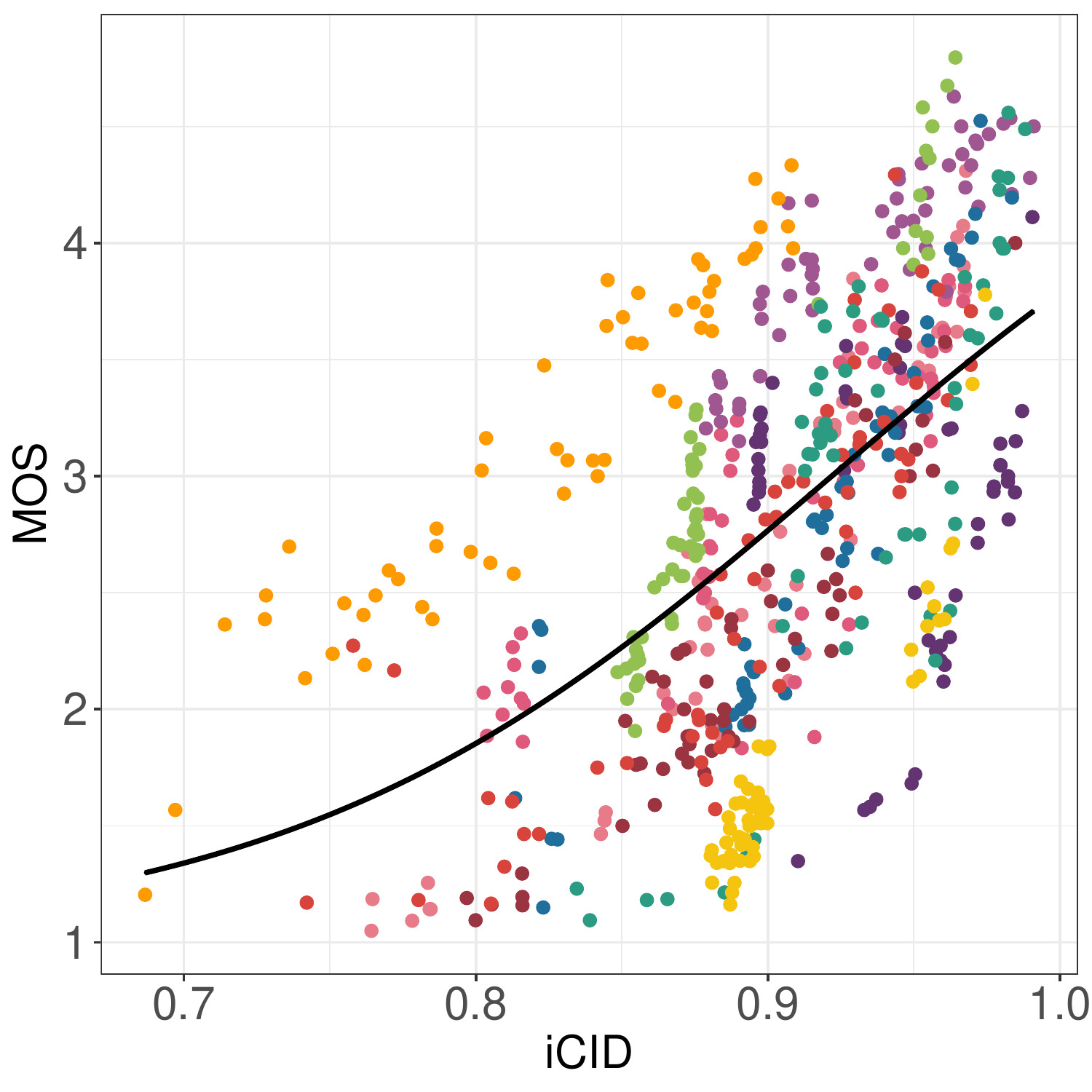}
  \end{subfigure}\hfill
  \begin{subfigure}[h!tb]{0.5\linewidth}
  \centering
    \includegraphics[width=0.23\linewidth]{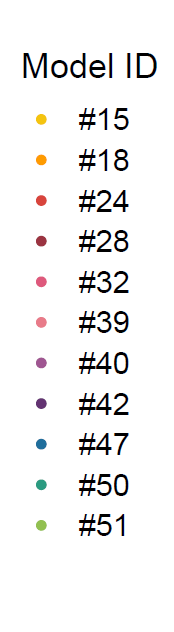}
  \end{subfigure}
 \caption{MOS vs.  quality metric values for the test set of textured meshes.  Each point represents a distorted stimulus identified by its source model. The curve shows the logistic regression.}
  \label{fig:MOsvsMetrics}
\end{figure}

\modified{Figure \ref{fig:Perf_testset} summarizes the performance of \textit{Graphics-LPIPS} and compares it to state-of-the-art Image Quality Metrics (IQMs), including the original LPIPS, on the test set of each of our five folds (each fold containing around 600 stimuli obtained from 11 source models). We report the mean and standard deviation over the five folds, the results on each individual fold can be found in supplemental material. Plots illustrating the subjective scores versus the metric values on a representative fold are shown in Figure \ref{fig:MOsvsMetrics}. As for our metric, the IQMs were  computed on the snapshots taken from the main viewpoint of the stimuli. The parameters of the IQMs are provided in the supplementary material.}

We evaluated the performance of the metrics in terms of correlations and classification abilities.
For correlation measures,  the Pearson Linear Correlation ($PLCC$) and the Spearman Rank Order Correlation ($SROCC$) were chosen.  To account for saturation effects associated with human senses, we computed PLCC after a logistic regression that establishes a nonlinear mapping between subjective and metric scores. 
For the classification ability analysis, we considered that proposed in \cite{Krasula2016} which consists of assessing (1) the ability of the metric to distinguish between significantly different and similar pairs of stimuli and (2) the ability of the metric to predict which stimulus is of better quality in a pair of stimuli.
This analysis takes into account the uncertainty of the subjective scores and considers the Area Under the Curve (AUC) as the performance measure.  We denote $AUC_{DS}$ and $AUC_{BW}$ respectively for the 2 analysis scenarios.

%
%
%
%

The proposed metric shows a much better classification ability and correlation with MOSs than IQMs. \modified{Statistical tests on the logistic regression residuals yield p-values $<<$ 0.0001}. \modified{Additionally, our metric exhibits more consistent results over the different folds (smaller standard deviation), showing that it generalizes better to a wide variety of images.}
The poor performance of the IQMs reflects the challenging aspect of our dataset. We believe that this is related to (1) the process of selecting the 3000 stimuli, which samples a lot of stimuli for which two quality metrics did not agree (see section \ref{subsec:StimuliSelection}) and to (2) the large variability of source models and distortion combinations (mixed distortions) present in this dataset.


\subsection{Validation on a dataset of 3D meshes with vertex colors}
We evaluated the performance of our metric on the dataset of meshes with vertex colors reported in \cite{Nehme2021TVCG}, to assess its robustness. 
This dataset is composed of 480 animated stimuli, generated from 5 source models corrupted by simplifications and compressions applied on geometry and color: uniform quantizations applied on either (1) geometry or (2) color, simplifications that take into account either the (3) geometry only or (4) both geometry and color. Each distortion was applied with 4 different strengths adjusted manually in order to cover the whole range of visual quality from imperceptible to high levels of impairment.
Each stimulus was displayed in 3 viewpoints and animated with 2 short \modified{rigid motions (rotations and zooms)}. The dataset was obtained through a subjective study in VR based on the DSIS method. 

As each stimulus in this dataset was rated in 3 different viewpoints, we computed Graphics-LPIPS on snapshots taken \modified{from} each viewpoint. We did not consider the influence of animations. Thus, for a given stimulus, we averaged its mean opinion scores (MOSs) over the two animations. The database used is therefore composed of 240 stimuli\remove{ (i.e. Graphics-LPIPS was computed on 240 snapshots)}.
We included the results of IQMs computed on these snapshots as well as the results of the Color Mesh Distortion Measure (CMDM),
which is a model-based quality metric for 3D meshes with colors attributes (i.e. works entirely on the mesh domain). It incorporates perceptually-relevant geometry and color features and is based on a data-driven approach \cite{Nehme2021TVCG}. CMDM was computed only over the visible parts of the 3D model (visible vertices) in each viewpoint.
\begin{table}[h!tb]
\begin{center}
\caption{Performance comparison of different metrics on a dataset of meshes with vertex colors. For metrics marked with a *, the values are reprinted from the supplementary material of \cite{Nehme2021TVCG}.}
\label{tab:Perf_VCDB_vp}
\resizebox{\columnwidth}{!}{
\begin{tabular}{l c c c c c}
\hline
& Graphics-LPIPS & CMDM\textsubscript{vis}* &SSIM* & HDR-VDP2* & iCID*\\
\hline
$PLCC$ & \textbf{0.89} &\textbf{0.89} & 0.79 & 0.81& 0.86\\
$SROCC$ & \textbf{0.88} & 0.87 & 0.8 & 0.83 & 0.87\\		
\hline
\end{tabular}
}
\end{center}
\end{table}

Although our metric was trained on a different dataset with different models and different distortions and even different color representation (textures and not vertex colors), its performance is comparable4to that of CMDM which was learned on this dataset.
This shows the good robustness of our metric and validates (1) that it did not just learn the distortions that are specific to our dataset and (2) its ability to differentiate and rank stimuli from different source models and different distortions.
Moreover, Table \ref{tab:Perf_VCDB_vp} also shows that our metric can be computed on different viewpoints of the 3D object (even if it is not necessarily the main viewpoint) and still provide good results in correlation with MOSs.

Furthermore, we noticed that IQMs exhibit poorer performance on our dataset of textured meshes than on the dataset of meshes with vertex colors (see Figure \ref{fig:Perf_testset} and Table \ref{tab:Perf_VCDB_vp}). 
This difference in IQMs performance between the two datasets illustrates once again the challenging aspect of our dataset.

\modified{\subsection{Robustness to changes in lighting and material}\label{sec-litmat}}

\begin{figure}[h!tb]
	\centering
	\includegraphics[width=\linewidth]{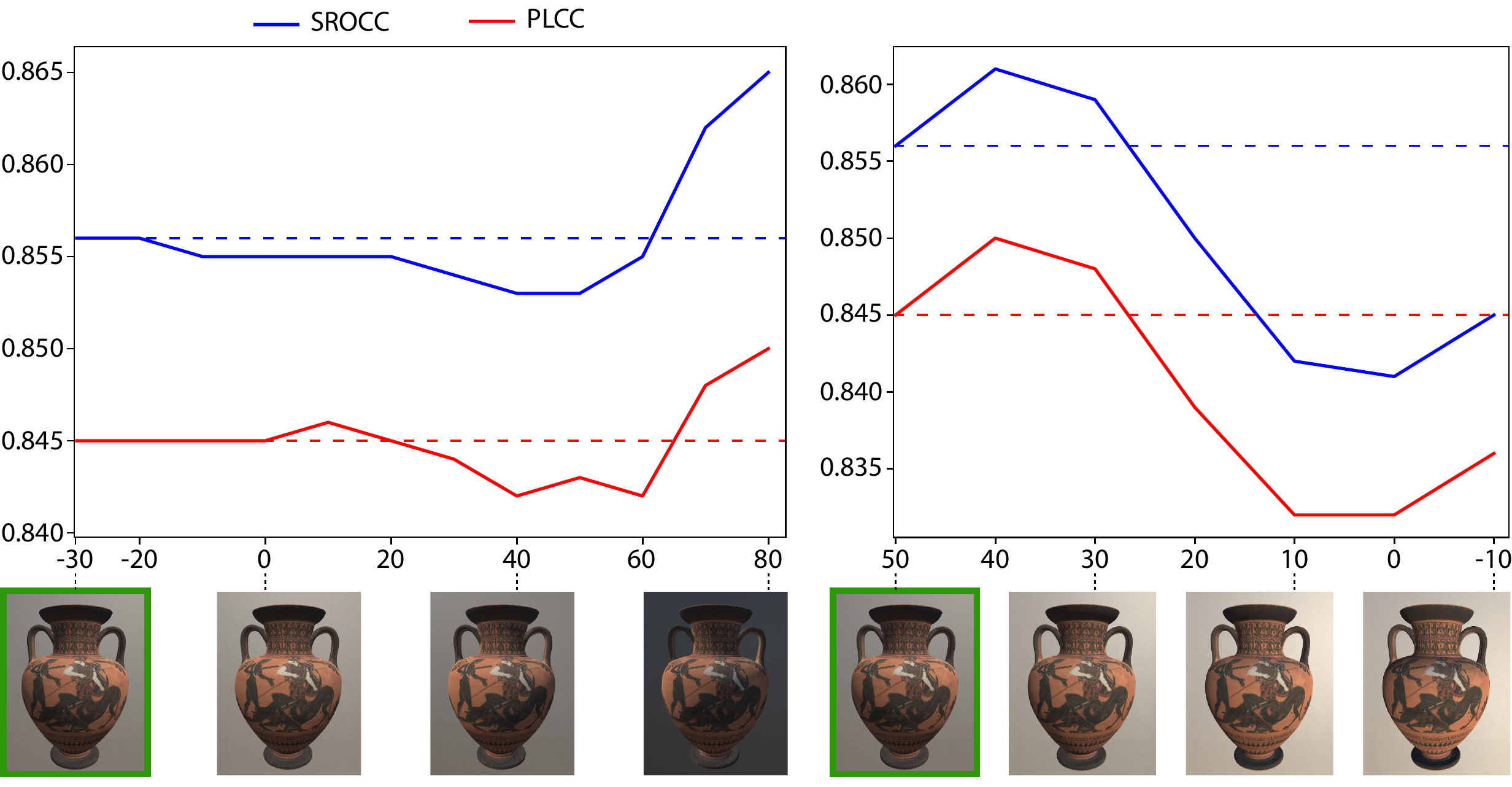}
	\caption{\modified{Performance of our metric when the light direction changes horizontally (left) or vertically (right). The angles in abscissa refer to the angles from the canonical axis (0 corresponds an horizontal line pointing straight to the front). The reference lighting lies on the left, its performances are depicted with a dash line. Samples of the rendered images are shown at the bottom, the reference lighting is circled in green.}}
	\label{fig:change_light}
\end{figure}

\begin{figure}[h!tb]
	\centering
	\begin{subfigure}[h!tb]{0.24\linewidth}
		\captionsetup{justification=centering,margin=-0.0cm}
		\includegraphics[width=\linewidth,trim=30 0 30 0, clip]{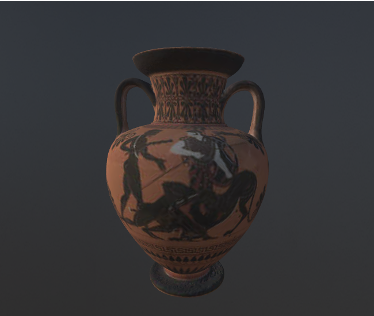}
		\caption{dimmed light \\ PLCC = 0.87\\ SROCC = 0.87} 
	\end{subfigure}
	\begin{subfigure}[h!tb]{0.24\linewidth}
		\captionsetup{justification=centering,margin=-0.0cm}	\includegraphics[width=\linewidth,trim=30 0 30 0, clip]{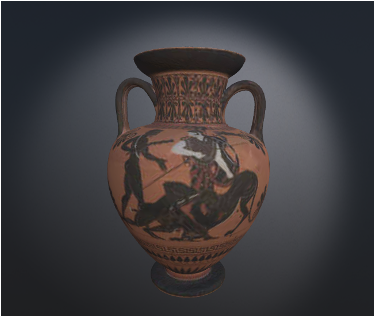}
		\caption{spotlight \\ PLCC = 0.81\\ SROCC = 0.80} 
	\end{subfigure}
	\begin{subfigure}[h!tb]{0.24\linewidth}
		\captionsetup{justification=centering,margin=-0.0cm}
		\includegraphics[width=\linewidth,trim=30 0 30 0, clip]{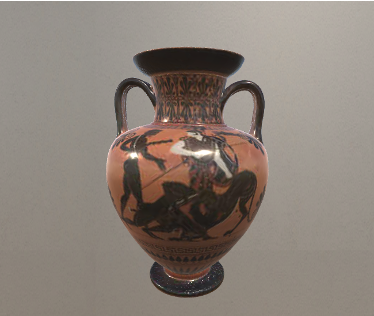}	
		\caption{glossy \\ PLCC = 0.72\\ SROCC = 0.72} 
	\end{subfigure}
	\begin{subfigure}[h!tb]{0.24\linewidth}
		\captionsetup{justification=centering,margin=-0.0cm}	\includegraphics[width=\linewidth,trim=30 0 30 0, clip]{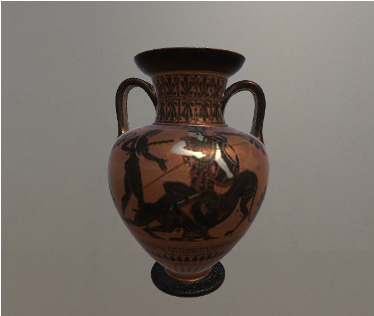}
		\caption{metallic \\ PLCC = 0.74\\ SROCC = 0.73} 
	\end{subfigure}
	\caption{\modified{Different conditions of lighting (dimmed light and spotlight) or material (glossy and metallic) along with their prediction scores (correlations).}}
	\label{fig:change_conditions}
\end{figure}

\modified{\textbf{Lighting.} One of the potential drawbacks of an image-based metric like ours is its potential dependency on the rendering conditions. Our metric is supposed to be able to evaluate the quality of a 3D model, regardless of how it is illuminated. To test whether our metric is robust to changes in the lighting conditions, we conducted the two following experiments: \\
\noindent (1) We moved the directional light toward the left side of the object (Figure \ref{fig:change_light}.left) and toward the bottom (Figure \ref{fig:change_light}.right). We then rendered the images with these new conditions and used the same network as before (\textit{without any retraining}) to compute results on the test set of our representative fold. Figure \ref{fig:change_light} provides the correlation results according to the angle from the canonical axis. For horizontal variations, the performance remains almost identical and even increases slightly for grazing angles. For vertical variations, the performance decreases slightly when the illumination becomes close to front lighting ($\approx0.01$ decrease in Pearson correlation). Given the fact that front lighting tends to completely mask geometric details, the observed robustness of our metric remains excellent, although the network never saw those lighting conditions before.\\
\noindent (2) We also tested two completely different lightings: one dimmed light coming from the left (Figure \ref{fig:change_conditions}.a) and a spot light coming from the front of the object (Figure \ref{fig:change_conditions}.b). In line with previous results, the first configuration leads to very good scores ($PLCC = 0.87$), showing even better performances than with the original illumination. The second condition leads to more degraded results ($PLCC = 0.81$). Still, although this configuration is completely different from the original one, its performances are still outperforming other IQMs.}

\modified{\textbf{Material.} Additionally, we evaluated how our metric behaves when changing the material property of an object. Note that, contrary to lighting conditions, our metric is not supposed to be robust against material change since the material intrinsically defines the appearance of the object itself. We rendered the objects with a glossy material (glossiness = $0.8$, metallic = $0$ in Unity PBR model, see Figure \ref{fig:change_conditions}.c) as well as with a metallic one (glossiness = $0.6$, metallic = $0.8$, see Figure \ref{fig:change_conditions}.d).  In that case, the performances of our metric decrease significantly with Pearson correlations of respectively $0.72$ and $0.74$. As mentioned above, this is expected since materials play a key role in the perception of an object, more specular material can lead to amplifying the visual effects of geometric deformation while minoring the texture ones. These results open interesting perspectives on the influence of the material in the perception of compression artifacts, that would require new subjective studies. }

\subsection{View-independent approach}
In order to have an automatic version of our metric that does not require manual selection of a main viewpoint for each 3D model, we considered another training scenario, using a set of snapshots of the model taken from different viewpoints. 
\modified{It may be especially relevant in our case }since all the stimuli in our dataset were animated with a full rotation (360 \textdegree) during the subjective test (see subsection \ref{sub:rendering}).
\modified{For each stimuli, we thus generated 4 snapshots taken from 4 camera positions regularly sampled on its bounding box and prepared the data as for the above network. We used the same training and testing set as our representative fold and used the same training parameters, randomly sampling $N_p=300$ patches from all possible viewpoints.} 
The results on the test set are reported in Table \ref{tab:Perf_testset_VPindependent}. 
For the IQMs, the global quality score of a stimulus is the average of the IQM values computed on its 4 snapshots.

\begin{table}[hptb]
  \begin{center}
    \caption{Performance comparison of different metrics on the test set of our textured mesh dataset, when several viewpoints are considered per stimulus.}
    \label{tab:Perf_testset_VPindependent}
		\resizebox{0.9\columnwidth}{!}{
    \begin{tabular}{l c c c c}
	  \hline
		 & Graphics-LPIPS & SSIM & HDR-VDP2 & iCID\\
		\hline
		$PLCC$ & \textbf{0.84} & 0.69 & 0.68& 0.68\\
		$SROCC$ & \textbf{0.83} & 0.67 & 0.69 & 0.67\\		
		$AUC_{DS}$ & \textbf{0.74} & 0.64 & 0.64 & 0.63\\		
		$AUC_{BW}$ & \textbf{0.96} & 0.9 & 0.88 & 0.88\\		
		\hline
    \end{tabular}
		}
  \end{center}
\end{table}

Comparing the results of Figure \ref{fig:Perf_testset} and Table \ref{tab:Perf_testset_VPindependent}, we observe that 
the performance of our metric decreases slightly when considering a view-independent approach.
This indicates that our manual selection of the main viewpoint for the source models is indeed relevant and helps the network. It also indicates that the perceptual pooling is not uniform: some parts or viewpoints of the objects have a stronger influence on the overall quality perceived by the observer. This effect depends on the metrics, as the performance of SSIM and iCID improves when considering multiple viewpoints per stimulus, while that of HDR-VDP remains stable.

\section{Application} \label{sec:applicationGraphics-LPIPS}
This section presents an application of the proposed metric and dataset. We used Graphics-LPIPS to annotate our whole dataset of textured meshes and study the influence of several factors on the visual quality.

Indeed, our subjective experiment involved only 3000 stimuli out of 343750 (i.e. only 3000 stimuli have a MOS value).  To annotate the remaining stimuli of the dataset, we applied Graphics-LPIPS to predict their MOS, referred to as pseudo-MOS. The pseudo-MOSs distribution of all stimuli of the dataset is provided in the supplementary material.

Annotating the entire dataset allowed us to explore the influence of the different compression parameters and their interactions on the subjective scores and thus on the perceived quality. We were also able to evaluate the impact of model characteristics on the perception of distortions. The subsequent subsections detail these analyzes.

\subsection{Influence of each distortion on perceived quality}
The perceptual quality of textured 3D content depends on both the geometry and color distortions.
Since the distortions in our dataset are of different natures (quantization, sub-sampling, simplification) and affect different aspects of the 3D  model (geometry and texture), we believe that their impact on the perceived quality is therefore very different.
In this subsection, we provide an in-depth analysis of the effect of each distortion on the perceived quality. We also determine which distortions affect the quality scores the most. To do so, we ran a Multivariate Analysis of Variance (ANOVA: $LoD_{simpL} \times qp \times qt \times T_{Q} \times T_{S}$) on the quality score of the entire dataset. \modified{All of the five distortions affect significantly the perceived quality (p-value $<<$ 0.0001), the full ANOVA table can be found in supplemental. }.\remove{The most significant results are presented below.}

\subsubsection{Influence of the geometry and texture coordinate quantization}\label{sub:qp-qt}

\begin{figure}[h!tb]
  \centering
  \begin{subfigure}[h!tb]{0.49\linewidth}
    \includegraphics[width=\linewidth]{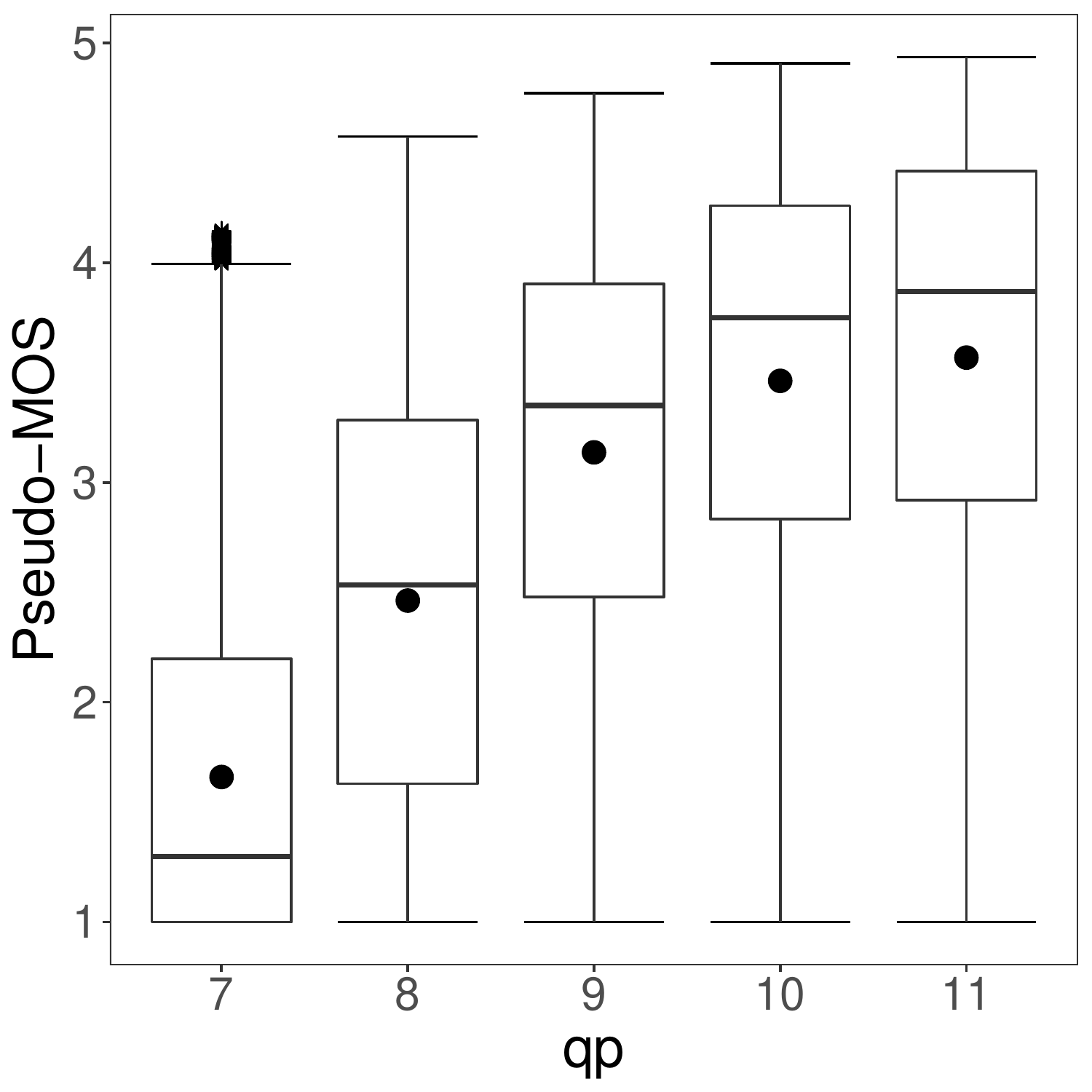}
    \caption{} 
    \label{fig:Influence_qp_qt.a}
  \end{subfigure}\hfill
	\begin{subfigure}[h!tb]{0.49\linewidth}
    \includegraphics[width=\linewidth]{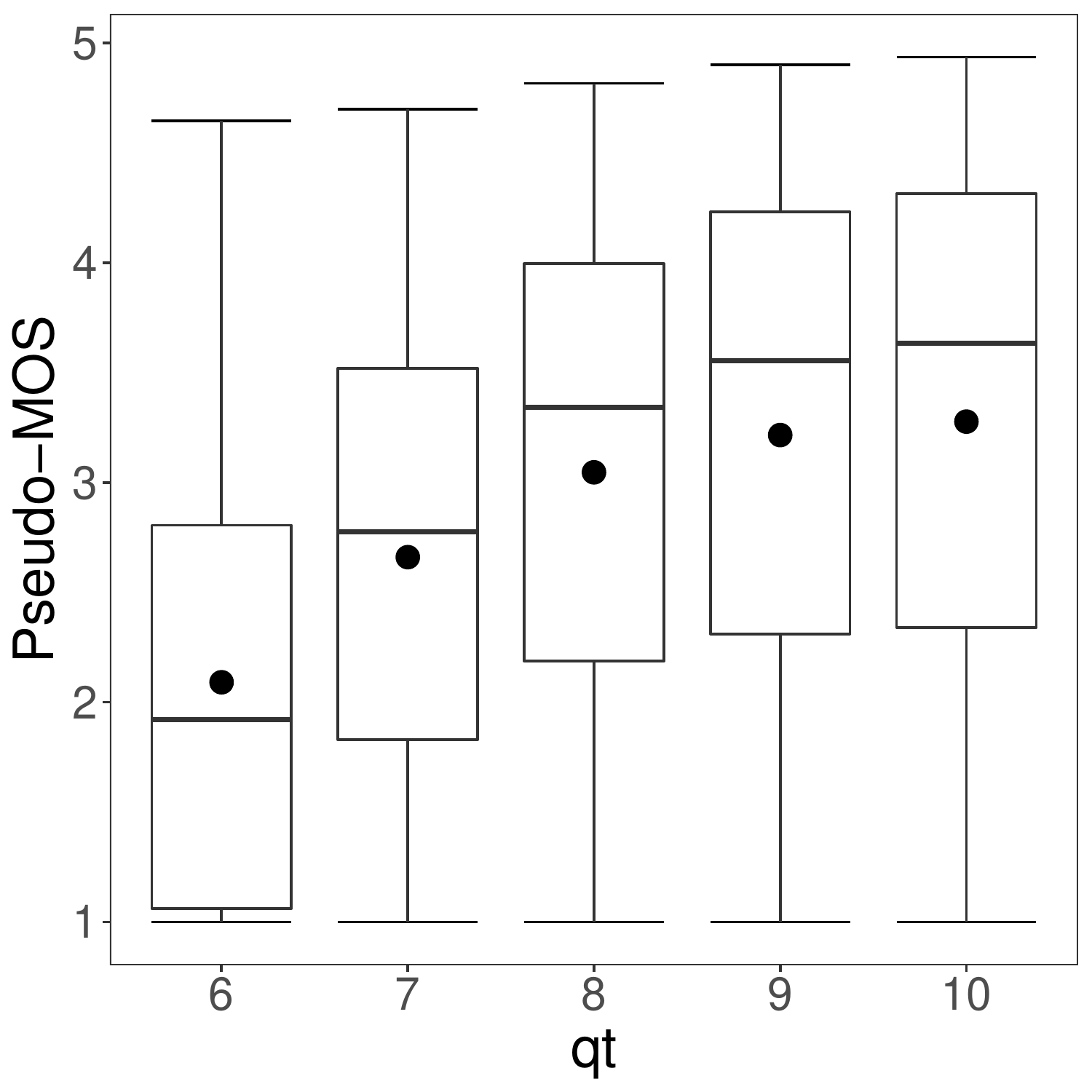}
    \caption{}
    \label{fig:Influence_qp_qt.b}
  \end{subfigure}
 \vspace{-1.5em}
 \caption{Boxplots of MOSs obtained for the quantization of the (a) vertices' positions $qp$ and (b) texture coordinates $qt$. Mean values are displayed as circles.}
  \label{fig:Influence_qp_qt}
\end{figure}

Figures \ref{fig:Influence_qp_qt.a} and \ref{fig:Influence_qp_qt.b} show the impact of the quantization parameters on the visual quality of 3D models.
As expected, quantizing the \modified{vertex} position or texture coordinates with too few bits can seriously degrade model quality. 
The advantage of using fewer quantization bits is the size reduction of the compressed files, however the resulting visual quality is vastly different from that of the original source model. 
Therefore, choosing the optimal/correct quantization parameters for an application depends on the intended quality as well as the size constraint. This is known as Rate-Distortion (RD) optimization.

\subsubsection{Influence of the LoD simplification}\label{sub:lod}
\begin{figure}[h!tb]
  \centering
  \begin{subfigure}[h!tb]{0.49\linewidth}
    \includegraphics[width=\linewidth]{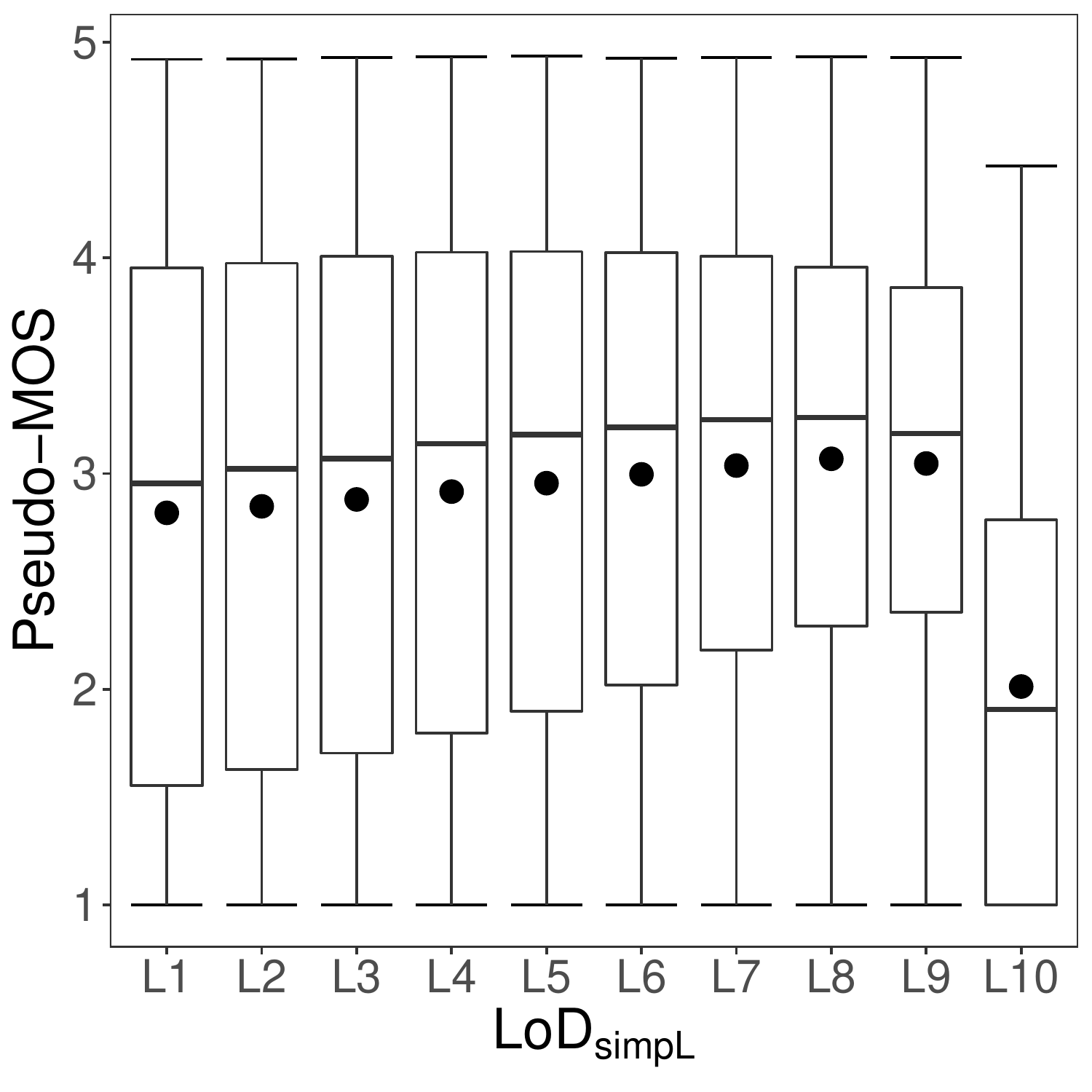}
     \caption{} 
     \label{fig:Influence_LOD.a}
  \end{subfigure}\hfill
	\begin{subfigure}[h!tb]{0.49\linewidth}
    \includegraphics[width=\linewidth]{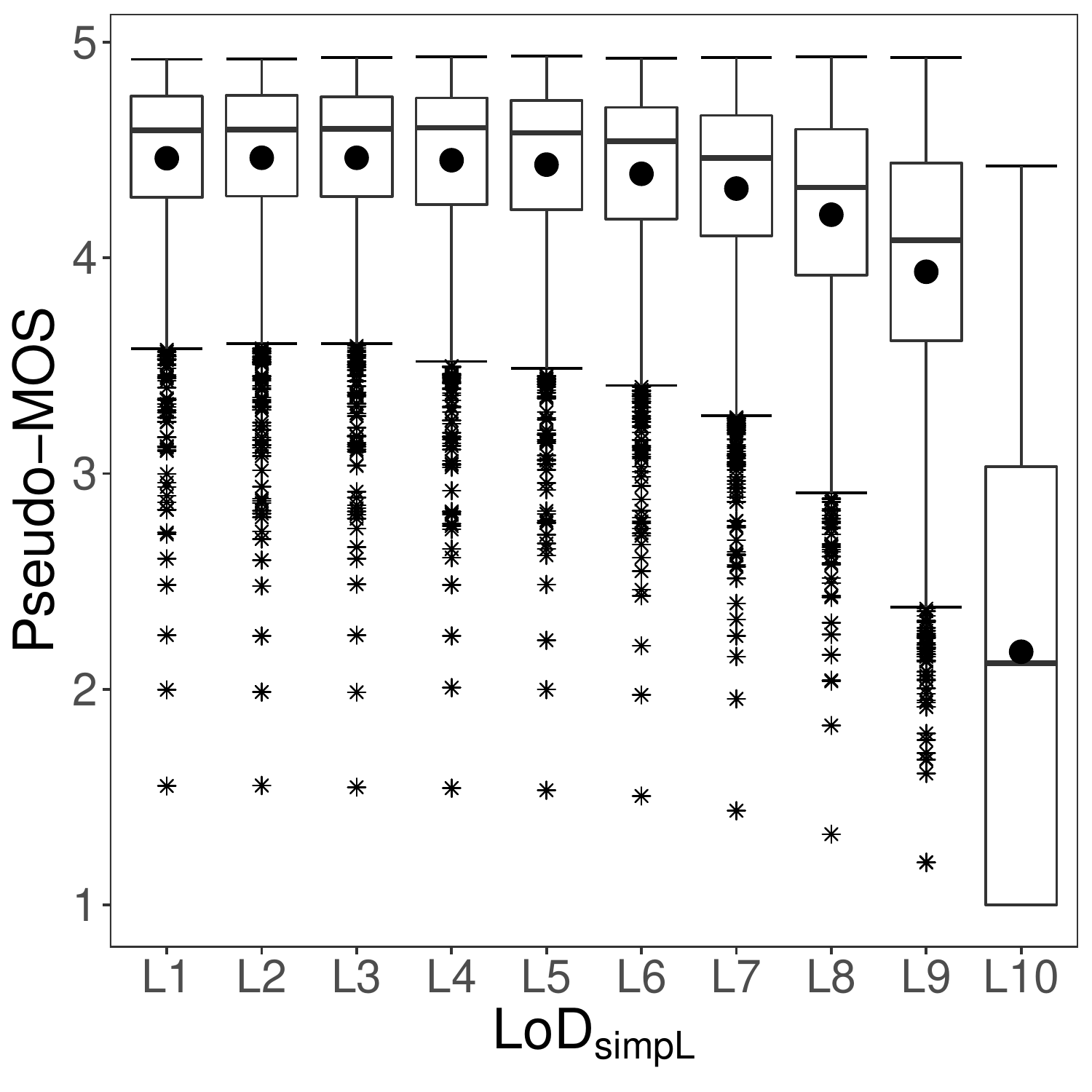}
    \caption{}
    \label{fig:Influence_LOD.b}
  \end{subfigure}
   \vspace{-1.5em}
 \caption{ Boxplots of MOSs obtained for the  LoD simplification $LoD_{simpL}$ (a) for all the stimuli and (b) for the least geometry quantized stimuli ($qp=11$ \& $qt=10$). Mean values are displayed as circles.}
  \label{fig:Influence_LOD}
\end{figure}

When looking at Figure \ref{fig:Influence_LOD.a}, it appears that the most simplified stimuli ($L7$, $L8$, $L9$) rated slightly better than the less simplified stimuli ($L1$, $L2$, $L3$). This is counter-intuitive and \modified{could }led us to think that simplifying the models with high strength did not introduce markedly visible impairments. This is not strictly true: it is actually highly dependent on the geometry quantization level. In fact, if we consider only the subset of the least geometry quantized stimuli ($qp=11$ \& $qt=10$), we see that the MOS logically decreases as the simplification level increases (see Figure \ref{fig:Influence_LOD.b}). There is thus a significant interaction between the geometry quantization of the model and its levels of detail (p-value $<<$ 0.0001). Subsection \ref{sub:lod:qp} details this point.\\
Note that for the most simplified level $L10$, it is a bit peculiar: for $L10$, the models are brutally/roughly simplified to have about 2000 faces. This is very degrading (regardless of the $qp$ and $qt$ values), especially for dense models with the highest number of vertices.

\subsubsection{Influence of the texture compression and sub-sampling}\label{sub:ts-tq}
\begin{figure}[h!tb]
  \centering
  \begin{subfigure}[h!tb]{0.49\linewidth}
    \includegraphics[width=\linewidth]{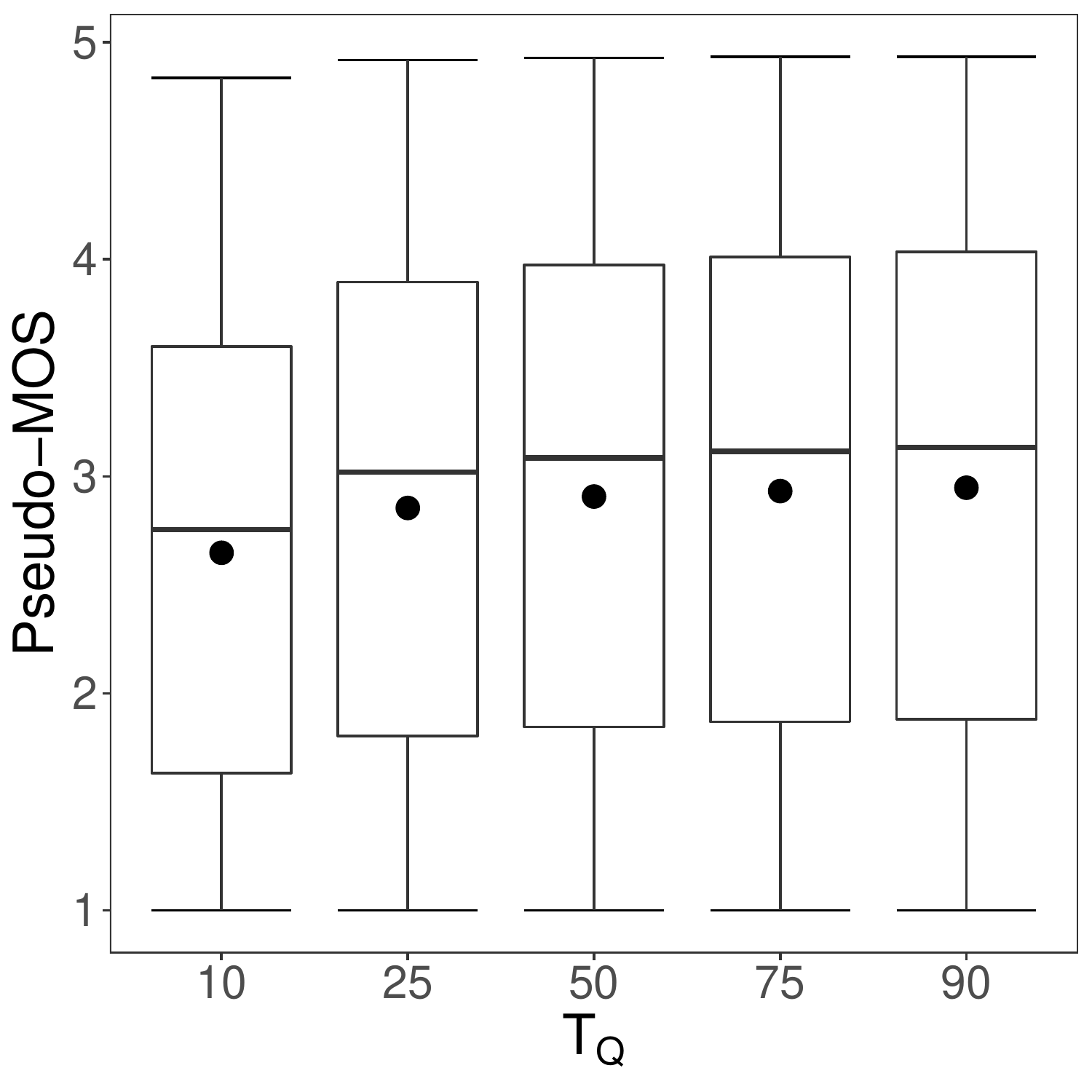}
	\caption{}
	\label{fig:Influence_Tq_Ts.a}
  \end{subfigure}\hfill
	\begin{subfigure}[h!tb]{0.49\linewidth}
    \includegraphics[width=\linewidth]{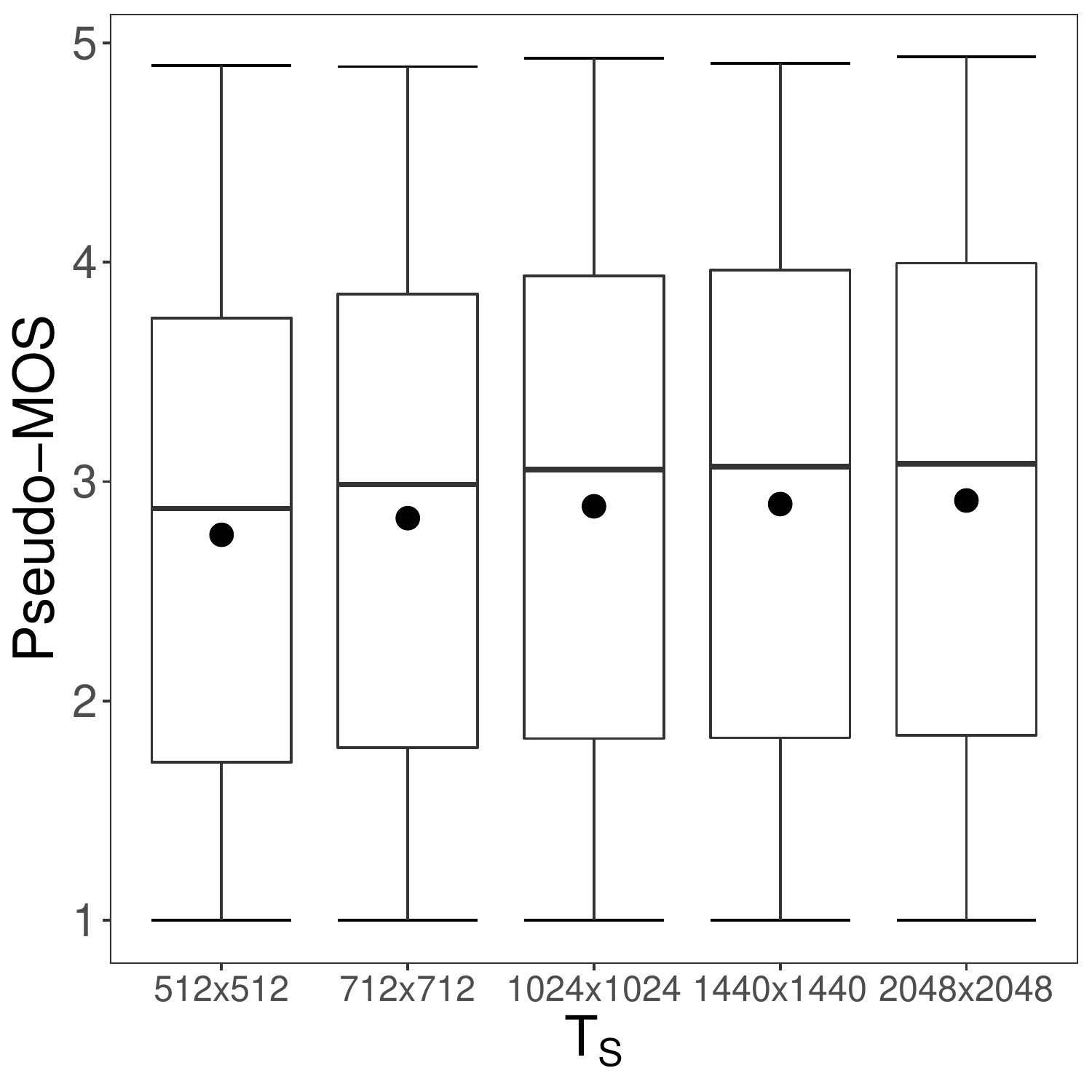}
	\caption{}
     \label{fig:Influence_Tq_Ts.b}
  \end{subfigure}
   \vspace{-1.5em}
 \caption{Boxplots of MOSs obtained for the texture (a) compression $T_Q$ and (b) sub-sampling $T_S$. Mean values are displayed as circles.}
  \label{fig:Influence_Tq_Ts}
\end{figure}

Figures \ref{fig:Influence_Tq_Ts.a} and \ref{fig:Influence_Tq_Ts.b} show the impact of the two distortions applied to the texture map (JPEG compression $T_Q$ and sub-sampling $T_S$) on the perceived quality. While their effect is statistically significant, the impact of these distortions on the MOS is not \modified{as} obvious as that of texture coordinates quantization \modified{(shown in Figure \ref{fig:Influence_qp_qt.b})}.
Figure \ref{fig:Influence_Tq_Ts.a} shows that for $T_Q \geq 50$, the increase of the texture quality does not seem to affect the overall perceived quality.\\
\modified{For texture sub-sampling, Figure \ref{fig:Influence_Tq_Ts.b} shows that increasing the texture resolution $T_S$ more than $712 \times 712$ did not overall influence the perceived quality. This may seem logical since the resolution of the stimulus videos shown in the experiment was $650 \times 650$. However, we recall that texel density depends on the surface area and even though render resolution is smaller than the texture resolution it does not mean that distortions will always be imperceptible (particularly when the UV mapping is non-uniform).}
 The impact of texture sub-sampling $T_S$ is emphasized when considering its interaction with texture compression $T_Q$ (see subsection \ref{sub:ts:tq}).\\
Thus for our visualization conditions, it seems that we can push the JPEG compression level and sub-sample the texture heavily without impacting the overall quality of the stimulus. This allows to drastically reduce the size of the compressed data.

\subsection{Influence of distortion interactions on perceived quality}
Based on the results of the previous subsection, we believe that the impact of the combinations of the different types of distortions differ from the cumulative impact of each distortion applied alone. 
This subsection presents the distortion interactions that have the most impact on the perceived quality of textured meshes. Other interesting interactions are provided in the supplementary material.

\subsubsection{Interaction of LoD simplification and position quantization}\label{sub:lod:qp}
\begin{figure}[hptb]
    \centering
	\includegraphics[width=\linewidth]{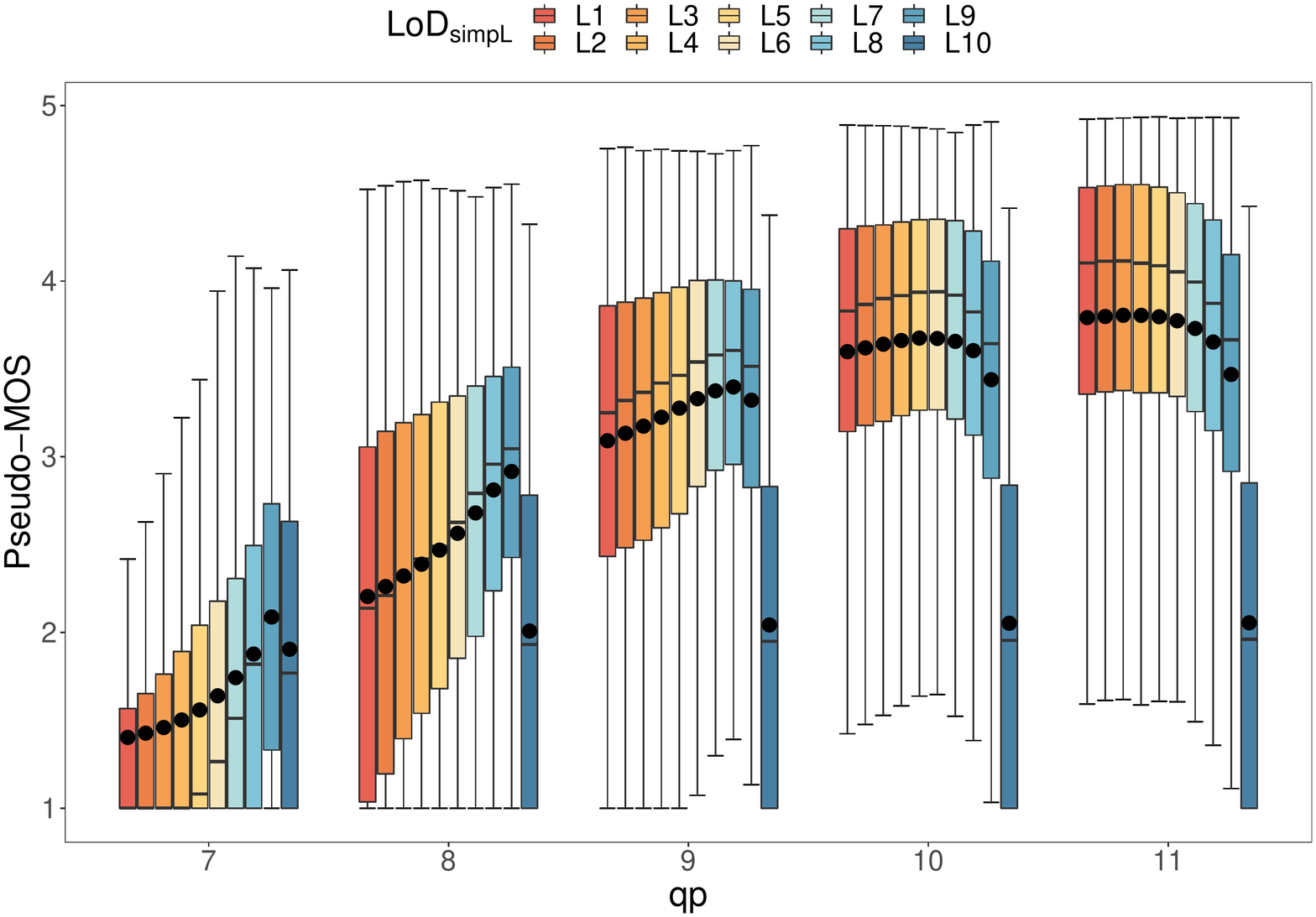}
	\caption{Boxplots of MOSs illustrating the interaction between the LoD simplification $LoD_{simpL}$ and the quantization of the model’s positions $qp$.}
	\label{fig:LOD:qp}
\end{figure}

The perception of geometry quantization artifacts depends strongly on the level of details of the stimulus (significant interaction with a p-value $<<$ 0.0001). 
Figure \ref{fig:LOD:qp} illustrates this interaction.
\begin{figure}[h!ptb]
	
  \begin{subfigure}[h!tb]{0.325\linewidth}
  \centering
	  \captionsetup{justification=centering}
    \includegraphics[width=0.9\linewidth]{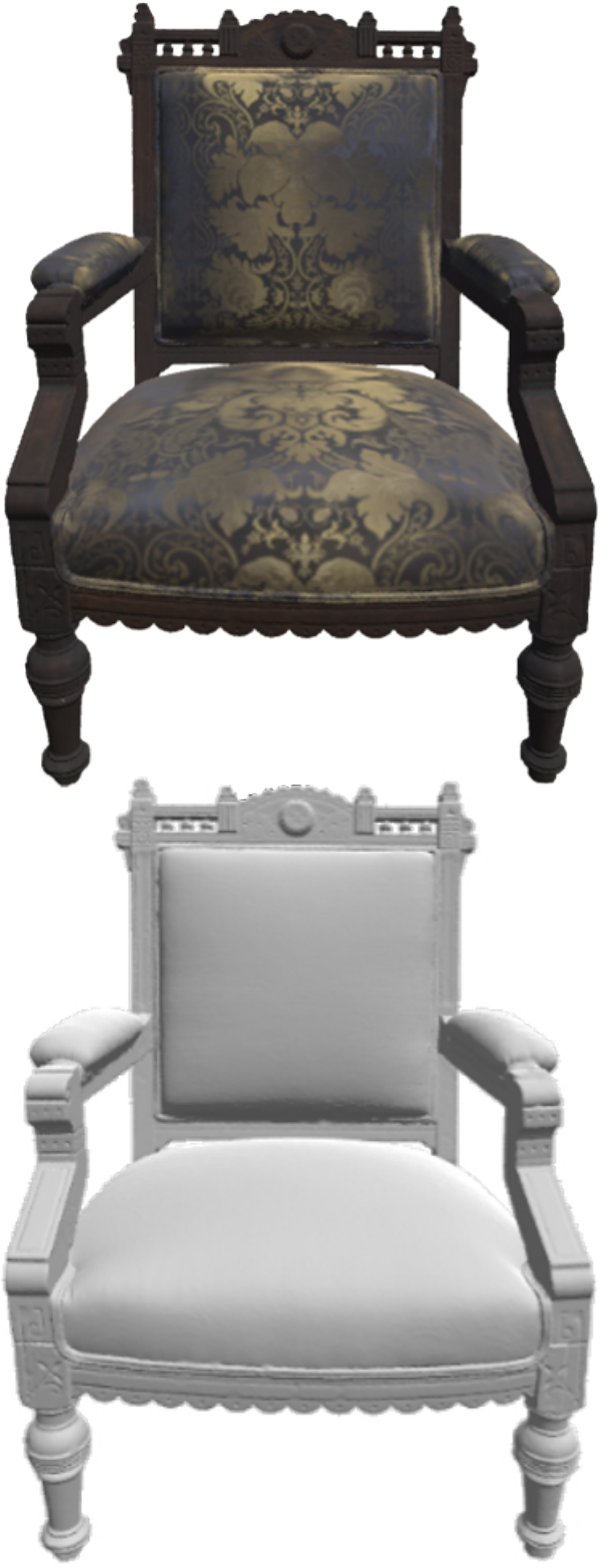}
    \caption{\\Source model\\Model \#3}
     \label{fig:LOD:qp_Exp.a}
  \end{subfigure}
	\begin{subfigure}[h!tb]{0.325\linewidth}
		\centering
	  \captionsetup{justification=centering}
    \includegraphics[width=0.9\linewidth]{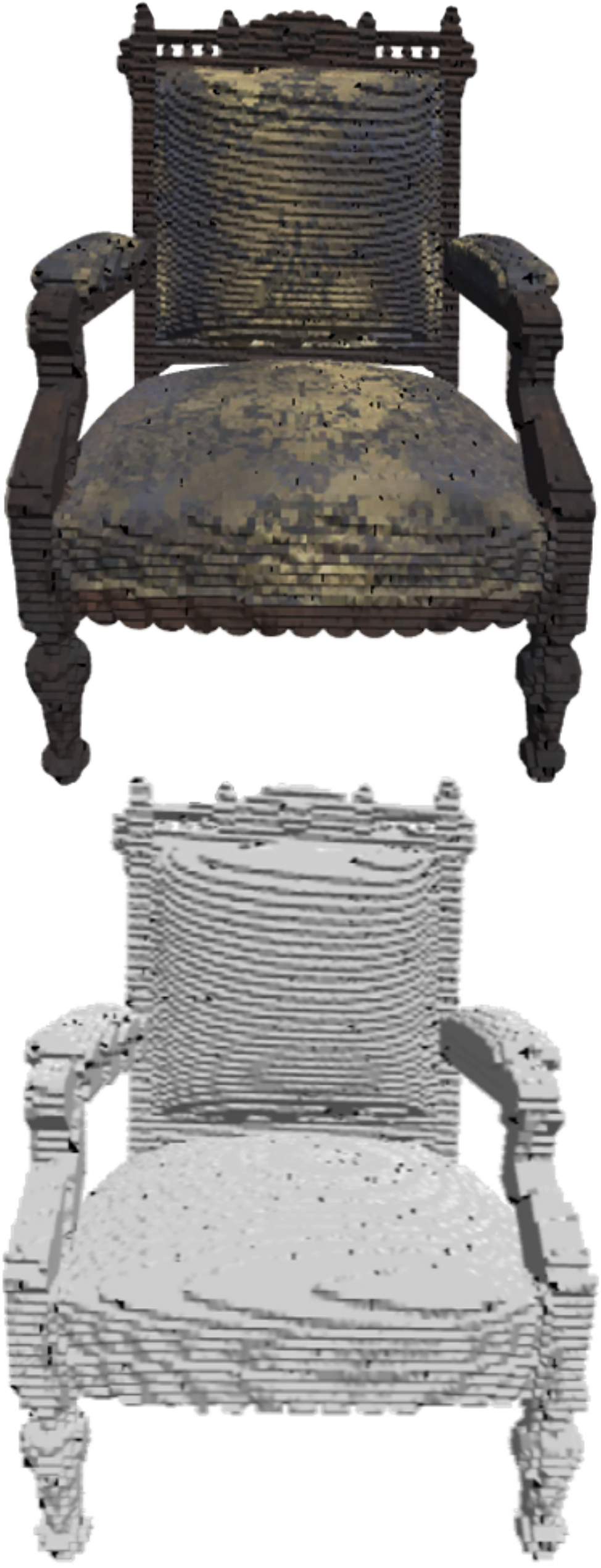}
    \caption{$\textbf{L1}| \textbf{7}| 10| 2048\times2048| 90$\\Pseudo-MOS = 1}
    \label{fig:LOD:qp_Exp.b}
  \end{subfigure}
	\begin{subfigure}[h!tb]{0.325\linewidth}
		\centering
	  \captionsetup{justification=centering}
    \includegraphics[width=0.9\linewidth]{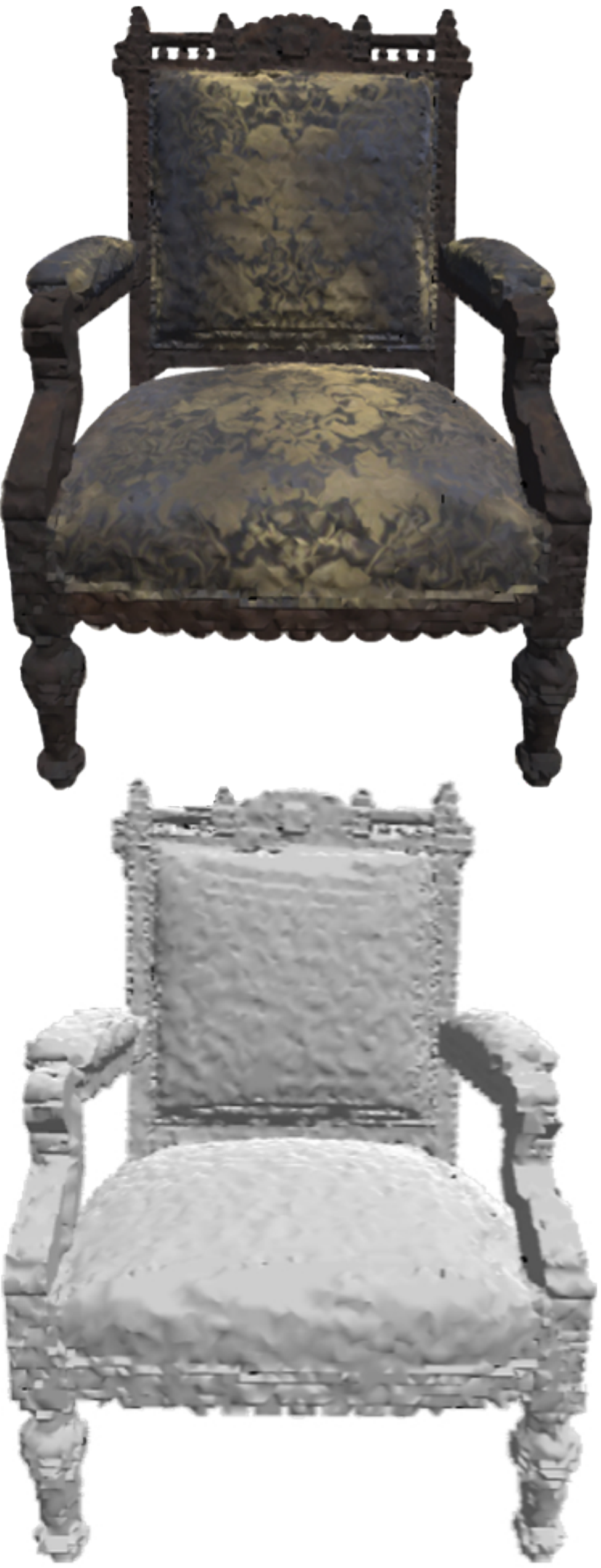}
    \caption{$\textbf{L9}| \textbf{7}| 10| 2048\times2048| 90$\\Pseudo-MOS = 2.4}
    \label{fig:LOD:qp_Exp.c}
  \end{subfigure}
 \caption{Visual example illustrating the interaction between the LoD simplification $LoD_{simpL}$ and the position quantization $qp$ regarding the perceived quality.  Acronyms refer to the following combination of distortion parameters: $LoD_{simpL}|qp|qt|T_S|T_Q$. The geometric quantization artifacts ($qp=7$) are more visible on the dense mesh (b) than on the simplified mesh (c).}
 \label{fig:LOD:qp_Exp}
\end{figure}

When quantizing stimulus's positions with too few bits ($qp \in \left\{7,\;8, \;9\right\}$), the MOS increases as the simplification level $LoD_{simpL}$ increases (i.e., the number of vertices decreases). 
The reason is that, the local geometry alteration (local contrast alteration) caused by a strong quantization is more visible on dense meshes (less simplified, $LoD_{simpL}=L1$) than on coarse meshes ($LoD_{simpL}=L9$). 
Figure \ref{fig:LOD:qp_Exp} illustrates a visual example in which we can see how the effect of the quantization of the vertex positions is much more visible on the dense model ($LoD_{simpL} = L1$). This is due to the fact that the frequency of artifacts created by geometry quantization is higher on a dense mesh than on a simplified mesh; this is what makes the artifacts more visible.

\vspace{1.5em}
\subsubsection{Interaction of the texture compression and sub-sampling}\label{sub:ts:tq}
\begin{figure}[hptb]

    \centering
	\includegraphics[width=\linewidth]{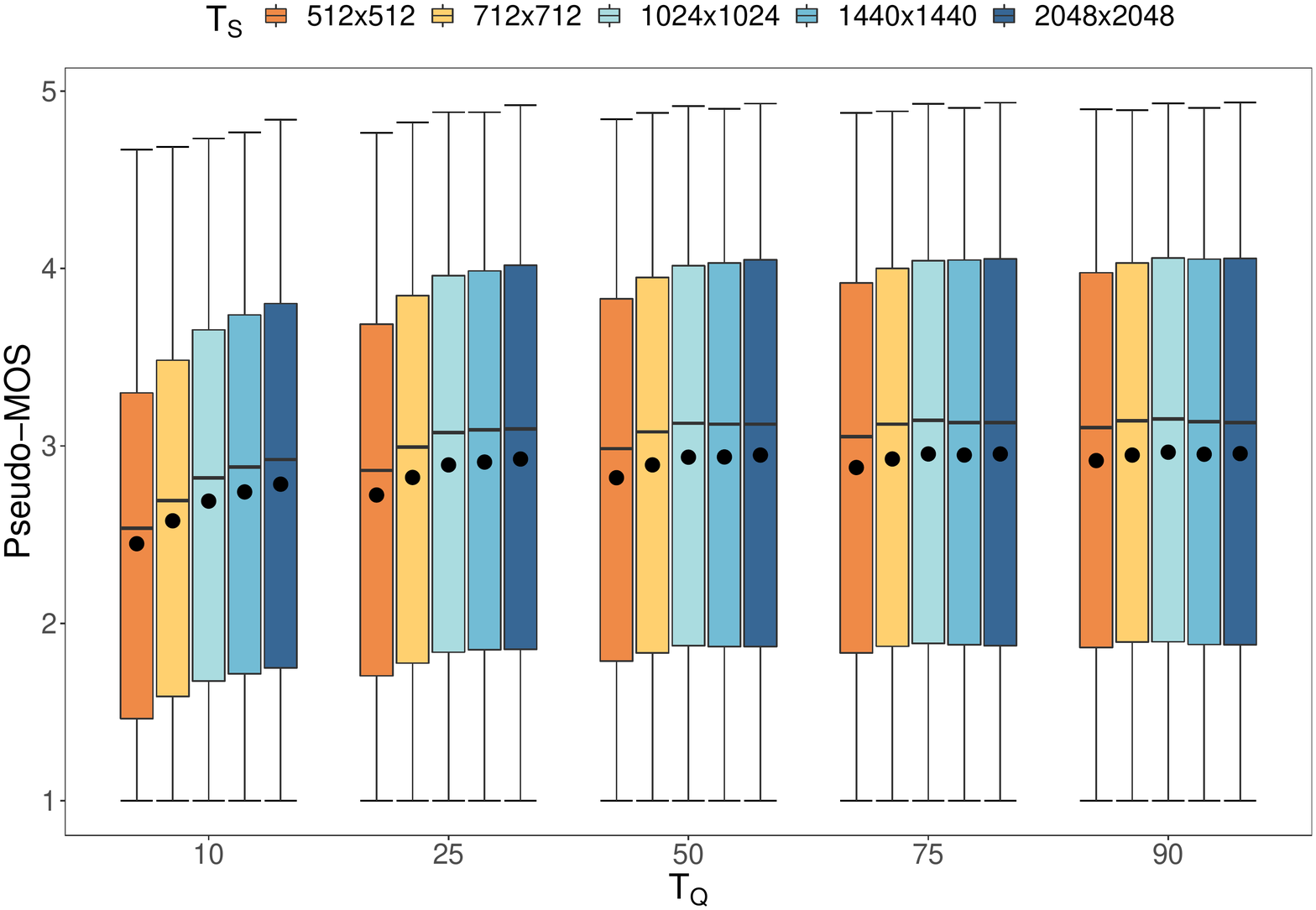}
	\caption{Boxplots of MOSs illustrating the interaction between the texture compression $T_Q$ and sub-sampling $T_S$.}
	\label{fig:Ts:Tq}
\end{figure}

There is a significant interaction (p-value $<<$ 0.0001) between the compression and sub-sampling applied on texture images. Figure \ref{fig:Ts:Tq} shows its impact on the perceived quality. For the lowest texture quality ($T_Q=10$), the MOS increases as the texture size $T_S$ increases. Overall, compression artifacts are less visible on larger textures. 
The reason is that the blocking artifacts caused by the JPEG compression are bigger/larger on screen for smaller textures.

We can also notice that stimuli with medium or low compressed textures ($T_Q \geq 50$) obtained almost the same MOSs regardless the texture size. This is coherent with what we observed in Figure \ref{fig:Influence_Tq_Ts.a}.

\modified{Those results show that, for the viewing conditions of our experiment, the perception of texture compression artifacts is subject to significant masking effects, probably due to the texture mapping, shading, and rasterization processes. Those masking effects makes the JPEG artifacts significantly less visible than for a \textit{natural} 2D image directly displayed on the screen.}

\subsection{Influence of content characteristics on perceived quality}
The content has a concealing effect on the perception of the distortions, which is consistent with the characteristics of the human visual system \cite{Karunasekera1995}.  Indeed,  for the same distortion parameters, the perceived quality may not be the same depending on the models and their characteristics. 

In this section, we evaluate the influence of content characteristics on the perception of distortions and thus on quality. To do so, we use the content characterization measures we developed in Section \ref{sec:contentCharac} ($SI_{Geo}$ and $SI_{Col}$). We group our 55 models into 5 clusters based on their geometric and color complexity. 
Thus, the first cluster ``$SI_{Geo}1$" contains the first 11 models with the least complex geometry (lowest $SI_{Geo}$ values), while ``$SI_{Geo}5$" designates the 11 models with the most geometric details (highest $SI_{Geo}$ values). \modified{Similarly, }``$SI_{Col}1$" denotes the first 11 source models with the least color details while ``$SI_{Col}5$" refers to the models with the richest texture. Our clusters are well dispersed in the $SI_{Geo} / SI_{Col}$ plane. 
An histogram representation of Figure \ref{fig:ContentCharac_Results}.a., provided in the supplementary material, illustrates that.\\

\subsubsection{Influence of geometric and color complexity on the perception of position quantization}\label{sub:qp:ModelChar}
To evaluate the influence of the characteristics of a model on the perception of the degradations generated by the quantization of the position of its vertices $qp$, we considered the subset of stimuli having a strongly quantized geometry and the levels of all other distortions fixed at their best levels (giving the best quality), i.e. we considered the stimuli having: $qp \in \left\{7,\;8, \;9\right\}$ \& $LoD_{simpL} \in \left\{L1,\;L2,\; L3\right\}$ \& $qt \in \left\{9,\;10\right\}$ \& $T_Q \in \left\{75,\; 90\right\}$ \& $T_S \in \left\{1440 \times 1440,\; 2048 \times 2048\right\}$. 

\begin{figure}[hptb]
\hspace{-1.5em}
  \begin{subfigure}[h!tb]{0.49\linewidth}
    \centering
	\includegraphics[width=1.1\linewidth]{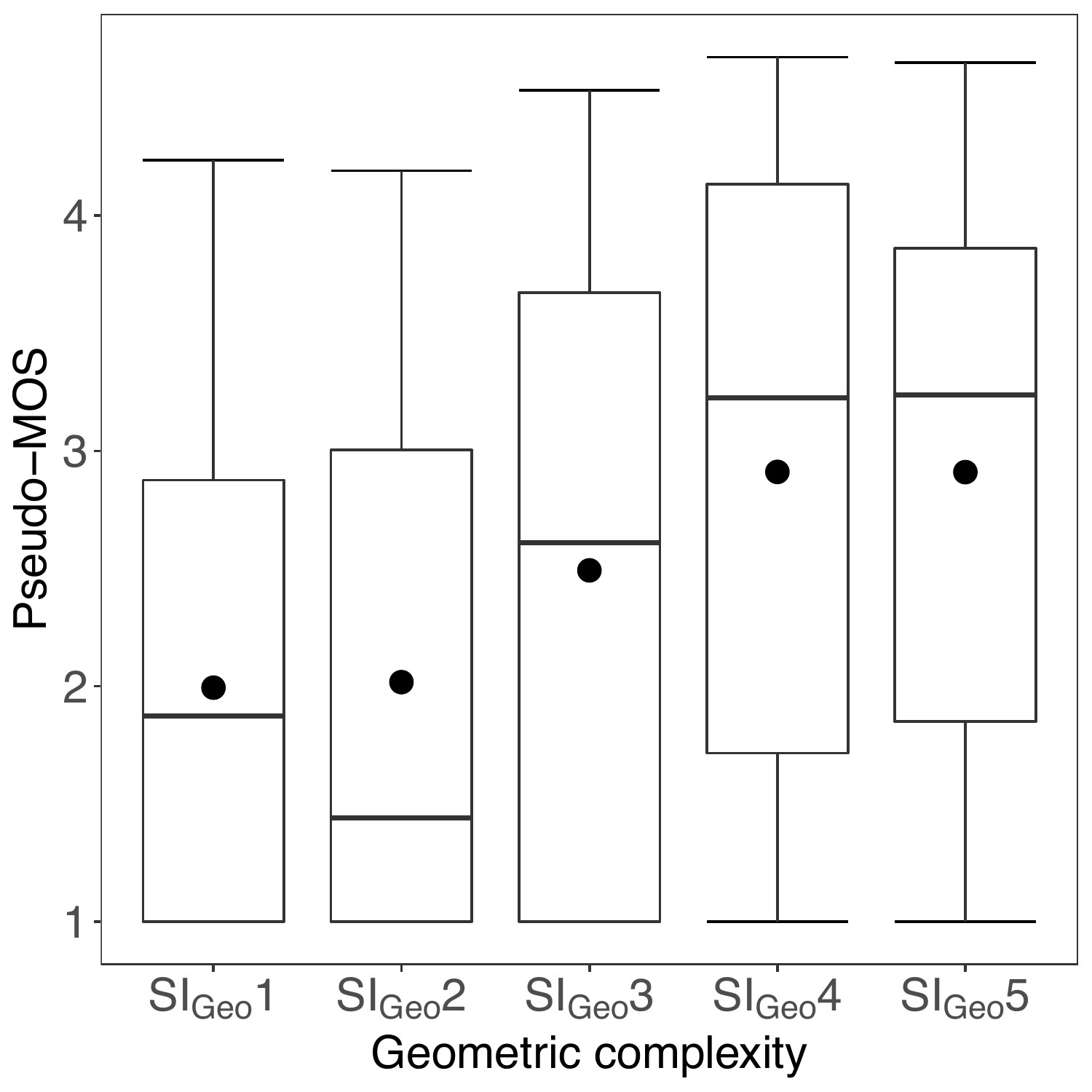}
	\caption{}
	\label{fig:qp:SIgeo}
\end{subfigure}
\hfill
  \begin{subfigure}[h!tb]{0.49\linewidth}
    \centering
	\includegraphics[width=1.1\linewidth]{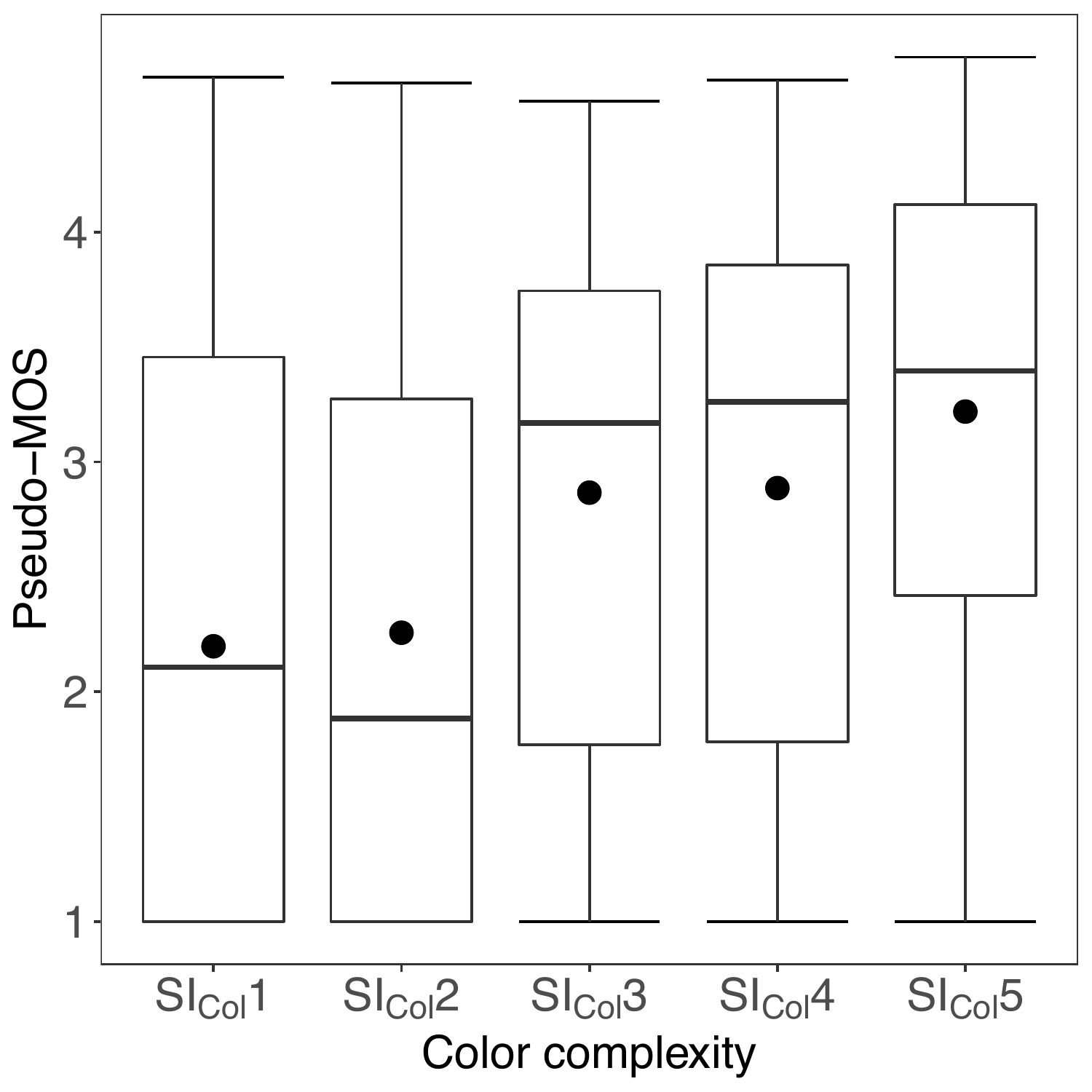}
	\caption{}
	\label{fig:qp:SIcol}
\end{subfigure}
\caption{Boxplots of the MOSs illustrating the influence of (a) geometric complexity $SI_{Geo}$ and (b) color complexity $SI_{Col}$ of the models on the perceived degradation of geometric quantization $qp$.}
\end{figure}

To assess the impact of geometry complexity, we eliminated the stimuli with rich textures ($SI_{Col}4$ and $SI_{Col}5$) in order to dissociate the influence of geometry and color and to avoid a possible masking effect of one on the other. According to ANOVA, a significant interaction exists between the geometric complexity of the model and the visual impact of the position quantization (p-value $<<$ 0.0001). Figure \ref{fig:qp:SIgeo} shows that the geometric information can masks the geometry alteration caused by the quantization of the vertices position: For the same quantization level $qp$, meshes with complex geometry ($\in \left\{SI_{Geo}4,\; SI_{Geo}5\right\}$) obtained higher MOSs than those with less complex geometry ($\in \left\{SI_{Geo}1,\; SI_{Geo}2\right\}$). 

Regarding the impact of color complexity, the results presented in Figure \ref{fig:qp:SIcol} show that for the same level of quantization, models with rich texture \modified{($\in \left\{SI_{Col}4,\; SI_{Col}5\right\}$)}
were judged to be of higher quality than those having simpler texture\modified{ ($\in \left\{SI_{Col}1,\; SI_{Col}2\right\}$)}
. These results corroborate those observed for point clouds, reported in \cite{QiLiu2021}. 

Thus, we can say that both geometry and color mask the geometric degradations of a quantized 3D model.

\subsubsection{Influence of geometric and color complexity on the perception of texture coordinates quantization}\label{sub:qt:ModelChar}
Following the same approach described in \ref{sub:qp:ModelChar}, we varied $qt  \in \left\{6,\;7,\;8\right\}$ and set the levels of all other distortions at their best levels ($LoD_{simpL} \in \left\{L1,\;L2,\;L3\right\}$ \& $qp \in \left\{10,\;11\right\}$ \& $T_Q \in \left\{75,\;90\right\}$ \& $T_S \in \left\{1440 \times 1440,\;2048 \times 2048\right\}$) in order to evaluate whether the model's geometric and color characteristics can mask the impairments caused by quantizing its UV map. 

\begin{figure}[hptb]
     \centering
 	\includegraphics[width=0.5\linewidth]{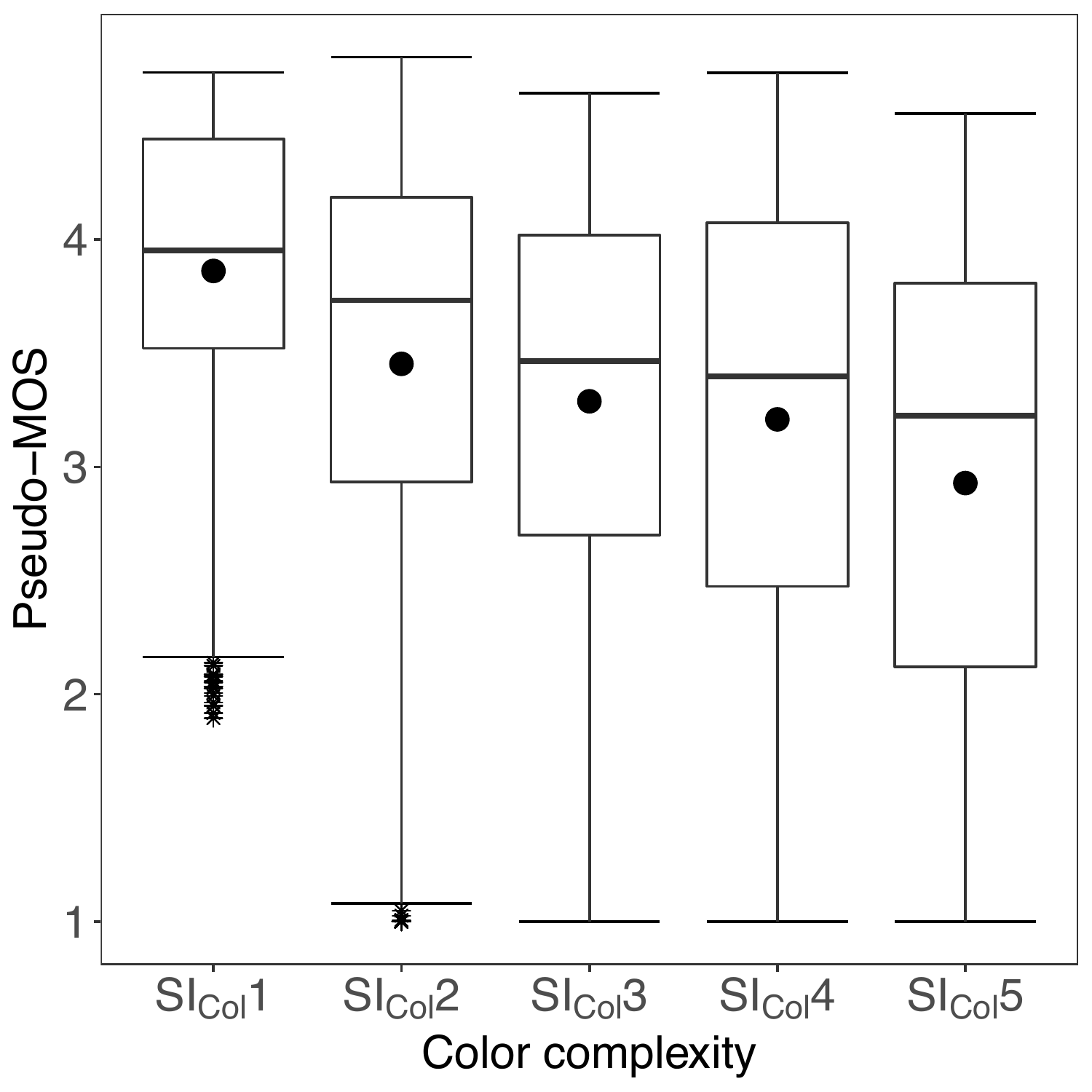}
 	\caption{Boxplots of the MOSs illustrating the influence of the color complexity $SI_{Col}$ of the models on the perceived degradation of texture coordinates quantization $qt$.}
 	\label{fig:qt:SIcol}
 \end{figure}

Figure \ref{fig:qt:SIcol} clearly shows that models with \modified{close-to-uniform textures (i.e., low contrast, with no high frequencies or noticeable structure, $\in \left\{SI_{Col}1,\; SI_{Col}2\right\}$)}
 are less sensitive to the UV map quantization than those with colorful and detail-rich textures \modified{($\in \left\{SI_{Col}4,\; SI_{Col}5\right\}$)}.

When analyzing the interaction between the geometric complexity $SI_{Geo}$ of the models and the quantization of their UV map $qt$, we realized that the influence of this interaction is more complex to evaluate,  yet it is significant (p-value $<<$ 0.0001).
The boxplots of the MOSs illustrating this interaction, as well as some visual examples are provided in the supplementary material.
We noticed that the impact of the UV map quantization on the visual quality depends not only on the geometric and color complexity of the model but also on the amount of texture seams (the level of fragmentation of its texture atlas): quantization artifacts are more visible on models exhibiting a large number of texture seams (texture atlas highly fragmented and/or not efficiently packed).
Further work is still needed to study the effect of texture seams.  
We speculate that the set of quality measures, reported in \cite{Maggiordomo2020}, to characterize the quality of the surface parametrization, notably the ``UV Occupancy” measure which assesses the quality of the atlas packing and the ``Atlas Crumbliness and Solidity” measure which captures the severity of texture seams in a given UV map, could be a good starting point.

\modified{\section{Limitations}\label{sec-lim}}

\modified{In the present work, the material information of each object is limited to one single diffuse texture, which is then mapped onto a Lambertian material for rendering. This pipeline thus does not integrate other texture maps representing physically-based material information like metalness and roughness, nor microgeometry information like normals or bumps. We made this technical choice for the sake of simplicity, given that this simple material representation remains reasonably realistic for many use cases and spans an already huge space of distortion parameters. As illustrated in Section \ref{sec-litmat}, the nature of the material (e.g., its specularity) obviously influences the visual impact of compression artifacts. In that respect, new user studies are needed to understand and model this influence.}

\modified{Our metric implementation makes the hypothesis that the perceived distortion of a global model can be modeled as an average of the local distortions from patches sampled either on a principal view or on multiple views. Even if we showed that this pooling provided better results than some other choices (L2, L3, or max pooling), it is highly probable that the reality of our visual system is quite different. Complex attention mechanisms must be at work, as raised by the results of our view-independent approach. We believe that one solution to capture this complexity would be to learn this pooling function, by letting the network learn where to focus its attention. }\modifiedbis{Similarly, the view-independent version of our metric is based on the automatic computation of 4 views sampled around the object. The integration of a data-driven view selection model (such as proposed in \cite{Secord:2011:PMO}) could certainly improve the results.} 

\modifiedbis{The proposed deep-learning metric has been shown to outperform standard state-of-the-art image quality metrics. More extensive comparisons could be conducted by including metrics dedicated to meshes and point clouds (e.g., \cite{Meynet2020a, Nehme2021TVCG})  as well as no-reference ones (e.g. \cite{Liu2021b}).}

\modifiedbis{The proposed dataset is large but based on only 55 source models. An extension could be considered by increasing the variety of source models (e.g., adding 300 sources) and limiting the number of distortions per source to keep subjective testing feasible.}

\section{Conclusion and future work}\label{sec:conclusion}
We produced a valuable large-scale textured meshes quality assessment dataset, with more than 343k distorted meshes derived from 55 source models corrupted by combinations of 5 real-world distortions, related to compression and simplification, applied on the geometry and texture. 
The source models cover a good diversity in visual contents. Indeed, we proposed three measures, based on spatial information and visual attention complexity, to quantitatively characterize the geometric, color and semantic complexity of the model. 
A subset of 3000 stimuli were rated in a crowdsourcing subjective experiment. This subset was selected to equitably cover the entire quality range, and to be challenging for objective quality metrics. 

Our dataset served us to develop a new image-based quality assessment metric for 3D graphics based on CNN. The metric, called Graphics-LPIPS, can be seen as an extension of LPIPS \cite{zhang2018}. It is computed on rendered snapshots of the 3D models. It employs a Siamese network fed with reference patches and distorted patches. We employed the AlexNet architecture with learning linear weights on top. The overall quality of the model is derived by averaging local patch qualities. The metric outperformed other image quality metrics in terms of correlations with subjective scores and classification abiltity on our textured mesh dataset.
Our metric also demonstrated a good robustness as it provided the best results on a dataset of meshes with vertex colors.

After validating the performance of Graphics-LPIPS, we used it to predict the quality scores of stimuli in our dataset not included in the subjective experiment. Annotating the entire dataset allowed us to analyze the influence of each distortion as well as that of their combinations on the perceived quality.  We also determine which distortions affect the quality scores the most.
We found a strong perceptual interaction between the geometry quantization of the mesh and its level of details. Indeed, quantization artifacts are less visible on coarse meshes.
Regarding the texture compression,  we showed that the quality level of the JPEG compression algorithm applied to the texture can be reduced to very low values while maintaining the quality of the final rendered stimulus.

Furthermore, we evaluated the influence of the geometry and color complexity on the perception of distortions. We observed that both color and geometry can mask the geometric degradations of a quantized 3D model. \modified{Models with close-to-uniform textures are less sensitive to UV map quantization}; however, the impact of this distortion on the visual quality depends also on the amount of texture seams.\\ 

\textbf{Further potential applications.}
Our dataset of 340K stimuli associated with pseudo-MOSs can be used to train no-reference quality metrics, but not only.
Another real-world use case/application of this dataset can be in the rate-distortion control and optimization.
This is possible because each of our stimuli is associated with a quality score and a file size, resulting from the compression methods we used for the source model and the texture. Thus, our dataset could allow to propose an analytical perceptual rate-distortion model capable of maximizing the visual quality of the reconstructed textured meshes subjected to a target bitrate. 

\textbf{Future works.}
\modified{As mentioned in Section \ref{sec-lim}}, some parts of the 3D objects probably have a stronger impact on the overall perceived quality than others. Therefore, we believe that an important improvement to our metric Graphics-LPIPS is to \modified{integrate and learn an attention model that would estimate the perceptual weight of each patch (from each view) on the global perceived quality. Still as mentioned in Section \ref{sec-lim}, a natural follow-up of our work is to replicate this study for 3D objects associated with more complex appearance models, e.g. represented by GGX parametrizations including normal,  diffuse, roughness, and specular maps.}\\

The source code of our metric and the datasets of textured meshes along with the subjective scores (individual quality scores, MOSs and Pseudo-MOSs) is publicly available online \footnote{https://github.com/MEPP-team/Graphics-LPIPS}.

\begin{acks}
This work was supported by French National Research Agency as part of ANR-PISCo project (ANR-17-CE33-0005).
\end{acks}

\bibliographystyle{ACM-Reference-Format}
\bibliography{References}

\end{document}